\documentclass[a4paper,11pt]{article}
\usepackage[utf8]{inputenc}
\usepackage{color}
\usepackage{cite}
\usepackage{hyperref}
\usepackage{amsmath}
\usepackage{graphicx}
\usepackage[top=0.4in, bottom=0.7in, left=0.60in, right=0.5in]{geometry}

\title{{\bf\Large{Analytic approach to calculations of mass spectra and decay constants of heavy-light quarkonia in the framework of Bethe-Salpeter equation}}}
\author{Eshete Gebrehana$^1$, Shashank Bhatnagar$^2$, Hluf Negash$^3$}
\begin{document}
\maketitle \small{$^1$Department of Physics, Addis Ababa
University, P.O.Box 1176 Addis Ababa, Ethiopia\\
$^2$Department of Physics, University Institute of Sciences, Chandigarh University, Mohali-140413, India\\
$^3$ Department of Physics, Samara University, Ethiopia}
\begin{abstract}
\normalsize{This work is an extension of the work in
\cite{bhatnagar18} to ground and excited states of $0^{++},
0^{-+}$, and $1^{--}$ of heavy-light ($c\overline{u},
c\overline{s}, b\overline{u}, b\overline{s}$, and $b\overline{c}$)
quarkonia in the framework of a QCD motivated Bethe-Salpeter
equation (BSE) by making use of the exact treatment of the spin
structure $(\gamma_{\mu}\bigotimes\gamma_{\mu})$ in the
interaction kernel, in contrast to the approximate treatment of
the same in our previous works \cite{hluf16, bhatnagar18}), which
is a substantial improvement over our previous works
\cite{hluf16,bhatnagar18}. In this $4\times 4$ BSE framework, the
coupled Salpeter equations for $Q\overline{q}$ (that are more
involved than the equal mass ($Q\overline{Q}$) mesons) are first
shown to decouple for the confining part of interaction, under
heavy-quark approximation, and analyically solved, and later the
one-gluon-exchange interaction is perturbatively incorporated,
leading to their mass spectral equations. The analytic forms of
wave functions obtained from these equations are then used for
calculation of leptonic decay constants of ground and excited
states of $0^{-+}$, and $1^{--}$ as a test of these wave functions
and the over all framework.}
\end{abstract}

\large{
\section{Introduction}
During the past few years, there is a growing interest in the
experimental and theoretical studies of heavy-light mesons. This
interest arose from the discovery of large $B_0-\overline{B}_0$
mixing, leading to the hope that CP violation in B-systems may be
observed. Further studies on heavy-light mesons are also important
for the determination of Cabibo-Kobayashi-Maskawa (CKM) mass
matrix elements. These studies on quarkonia need heavy quark
dynamics, which can provide a significant test of Quantum
Chromodynamics (QCD). Spectroscopy of heavy quarkonia have been
studied through non-perturbative QCD approaches, such as
NRQCD\cite{brambilla05}, QCD sum rule\cite{shifman79}, potential
models\cite{Godfrey85,Ebert03}, lattice QCD\cite{Bali97,Burch10},
Bethe-Salpeter equation (BSE)
method\cite{Mitra01,Alkofer01,Mitra92,Munczek93,bhatnagar14,bhatnagar06,bhatnagar11,glwang},
heavy quark effective theory\cite{neubert94}, Relativistic
Quantum Model (RQM)\cite{ebert11}, and Chiral perturbation
theory\cite{gasser84}.

It may be recalled that the discoveries of the low-lying
charmonium states and of open-charmed hadrons were instrumental
for the acceptance of quarks as truly dynamical entities in
general, and of the SM in particular. Thus studies of heavy
charmonium ($c\overline{c}$) and bottomonium ($b\overline{b}$)
states is a frontier area of research interest. Now, unequal quark
heavy meson, $c\overline{b}$ is the only bound state discovered
that comprises of two heavy quarks of different flavours, and acts
as an intermediate state between $c\overline{c}$, and
$b\overline{b}$ states. The discovery of $B_c$ state, has given a
new insight into heavy-quark dynamics, though its vector
counterpart, $B_c*$ has not yet been discovered in experiments.
The quark content of $B_c$ forbids its decays into two photons,
and can only decay through weak interactions, and have radiative
decays, and thus can lead to calculation of CKM matrix elements.
The same is true of other heavy-light mesons, $Q\overline{q} (q=
u,d,s)$.

The renewed interest in recent years in spectroscopy of these
heavy and heavy-light hadrons in charm and beauty sectors, which
was primarily due to experimental facilities the world over such
as BABAR, Belle, CLEO, DELPHI, BES etc.
\cite{Olive14,tanabashi2018,babar09,Ecklund08,belle10,cleo01},
have been providing accurate data on these hadrons with respect to
their masses and decays. In the process many new states have been
discovered such as $\chi_{b0}(3P), \chi_{c0}(2P), X(3915),
X(4260), X(4360), X(4430), X(4660)$ \cite{Olive14}. Further, there
are also open questions about the quantum number assignments of
some of these states such as $X(3915)$ (as to whether it is
$\chi_{c0}(2P)$ or $\chi_{c2}(2P)$ \cite{guo,olsen}).  Currently
there is a lot of excitement about $XYZ$ particles, that are new
charmonium like states such as, $Z_c(3900)$,  $Z_c(4020)$
/$Z_c(4025)$ \cite{ablikim}, which were discovered in BESIII, the
states, $Y(4260)$ \cite{aubert,yuan}(discovered at BABAR), and
$X(3872)$ \cite{choi} (discovered at Belle). These particles show
different features than the conventional charmonium states, and
might be good candidates for exotic states, and might even be
hybrid or tetra-quark states, or loosely bound charmonium
molecules, which is one of the predictions of QCD.

Thus charmonium like states offer us intriguing puzzles. However,
since the mass spectrum and the decays of all these bound states
can be tested experimentally, theoretical studies on them may
throw valuable insight about the heavy quark dynamics. Studies on
these hadrons is particularly important, as it throws light on the
long ranged $Q\overline{Q}$ and $Q\overline{q}$ potential, that
has not been derived from QCD so far.

In our works, we are not only interested in studying the mass
spectrum of hadrons, which no doubt is an important element to
study dynamics of hadrons, but also the hadronic wave functions
that play an important role in the calculation of decay constants,
form factors, structure functions etc. for $Q\overline{Q}$, and
$Q\overline{q}$ hadrons. This is due to the fact that so far, one
of the central difficulties in tests of QCD is lack of knowledge
of hadronic wave functions. These hadronic Bethe-Salpeter wave
functions calculated algebraically in this work can act as a
bridge between the long distance non-perturbative physics, and the
short distance perturbative physics. This is further due to the
fact that, though these quarkonium states appear to be simple,
however, their production mechanism is still not properly
understood. Thus the wave functions calculated analytically by us
can lead to studies on a number of processes involving
$Q\overline{Q}$, and $Q\overline{q}$ states. Though the
ground-state quarkonia have been shown over the past decade to be
rather well described in nonrelativistic QCD, and N.R. potential
models, the heavy-light mesons are much more complicated, where
one really tests the wide range of soft, hard and soft-collinear
scales. Our basic aim has been to develop a model using $4\times
4$ BSE that can explain both mass spectrum of $Q\overline{Q}$, and
$Q\overline{q}$ states as well as their decay widths through
various processes using the same set of input parameters that are
fixed from their mass spectrum.

In this context, in some of the recent works
\cite{elias13,hluf15,hluf16,hluf17,bhatnagar18}, we have been
involved in working on the mass spectrum and decay properties of
equal mass ground and excited states of scalar, pseudoscalar,
vector, and axial vector $Q\overline{Q}$ quarkonia in the
framework of a $4\times4$ BSE. These include their leptonic
decays, two-photon decays, single photon radiative decays and two
gluon decays of these charmonium ($c\overline{c}$) and bottomonium
($b\overline{b}$) states which have been extensively studied by us
in the formulation of Bethe-Salpeter equation. However, we had not
so far generalized this  $4\times 4$ representation for two-body
($Q\overline{Q}$) BS amplitude framework to incorporate unequal
mass dynamics, which we have now done in the present work.
However, the price we have to again pay is to analytically solve a
coupled set of equations for all quarkonia, which we have again
explicitly shown get decoupled, in spite of the fact that we have
used the full structure of the BS wave function,
$\psi(\widehat{q})$ in calculation of
$\gamma_{\mu}\psi(\widehat{q})\gamma_{\mu}$ on the right side of
the Salpeter equations. Due to these facts, the system of coupled
Salpeter equations encountered in the present work are much more
involved and complex than the ones encountered in equal mass
quarkonia in \cite{bhatnagar18}. We have explicitly shown that
they lead to mass spectral equations with analytical solutions for
both masses, as well as eigen functions for the ground and excited
states for $0^{++}, 0^{-+}$, and $1^{--}$ for heavy-light hadrons
with quark composition, $c\overline{u}, c\overline{s},
c\overline{b}, b\overline{u}$, and $b\overline{s}$ in an
approximate harmonic oscillator basis. We then perturbatively
incorporate the One-Gluon-Exchange (OGE), and solve the spectrum
of these states. We wish to mention that in unequal mass systems
such as $Q\overline{q}$, the quarks are not very close together,
and the confining interaction dominates over the OGE interactions
due to which the perturbative incorporation of OGE term is a good
approximation. The analytical forms of eigen functions for ground
and excited states obtained as analytic solutions of spectral
equations are then used to evaluate the decay constants and decay
widths for leptonic decays of these decays as a test of our framework. \\

The study of these mesons involves unequal mass kinematics. The
unequal mass kinematics (that also gives the partitioning of
internal momentum of hadron) used by us that rests on
Wightman-Garding (W-G) definitions of momenta between individual
quarks is relativistic, and has the advantage that $P.\hat q =0$
irrespective of whether the individual quarks are on-shell
($P.q=0$) or off-shell ($P.q\neq 0$).  This W-G partitioning of
momenta between individual quarks is a natural choice of momentum
partitioning, that allocates most of the internal momenta to
heavier quark, while a smaller part of momentum to lighter quark,
such that $\widehat{m}_1 + \widehat{m}_2=1$,  while for equal mass
mesons, the momentum is shared equally between the two quarks,
which is what one expects.

The main advantage of our approach in comparison to other BSE
approaches is that, we follow analytic methods of solutions for
heavy-light quarkonia (whose equations are much more involved than
$Q\overline{Q}$) , that provide a much deeper in sight into the
mass spectral problem, and are able to obtain the mass spectrum in
terms of the principal quantum number $N$, and also in the
process, we get algebraic forms of wave functions that are used
for calculations of various transition amplitudes and decay
constants of quarkonia, in contrast to the purely numerical
approaches followed by the other works. The correctness of our
analytic approach can be judged by the fact that the plots of our
wave functions (see \cite{bhatnagar18}) are very similar to the
plots of wave functions obtained by numerical approaches \cite{glwang}. \\

This paper is organized as follows: In section 2, we introduce the
formulation of the $4\times 4$ Bethe-Salpeter equation under the
covariant instantaneous ansatz, and derive the hadron-quark
vertex. In sections 3, 4, and 5, we derive the mass spectral
equation of heavy-light scalar, pseudoscalar, and vector mesons
respectively. In Sections 6, and 7, we derive the decay constants
$f_P$ for pseudoscalar, and $f_V$ for vector $Q\overline{q}$
states respectively. In section 8, we provide the numerical
results and discussion.

\section{Formulation of the 4$\times$ 4 Bethe-Salpeter equation}
We give here the main points about the 4$\times$ 4 BSE under the
Covariant Instantaneous Ansatz (CIA), which is a Lorentz-invariant
generalization of Instantaneous Approximation (IA), which is used
to derive the 3D Salpeter equations\cite{hluf16,bhatnagar18,yang}.
We start with a 4D BSE for quark- anti quark system with quarks of
constituent masses, $m_{1}$ and $m_{2}$, written in a $4\times 4$
representation of 4D BS wave function $\Psi(P,q)$ as:

\begin{equation}
S_{F}^{-1}(p_{1})\Psi(P,q)S_{F}^{-1}(-p_{2}) =
\frac{i}{(2\pi)^{4}}\int d^{4}q'K(q,q')\Psi(P,q')
\end{equation}

where $K(q,q')$ is the interaction kernel between the quark and
anti-quark, and $p_{1,2}$ are the momenta of the quark and
anti-quark, which are related to the internal 4-momentum $q$ and
total momentum $P$ of hadron of mass $M$ as, $ p_{1,2\mu} =
\hat{m}_{1,2}P_{\mu} \pm q_{\mu}$, where,
$\hat{m}_{1,2}=\frac{1}{2}[1\pm\frac{(m^{2}_{1}-m^{2}_{2})}{M^{2}}]$,
always satisfy, $\hat{m}_{1}+\hat{m}_{2}=1$, and is a natural
choice that allocates most of the momentum to the heavy quark,
while a smaller part of momentum to the lighter quark in a
heavy-light meson, but equal momenta to both quarks in
$c\overline{c}$ mesons.

Making use of Covariant Instantaneous Ansatz, where,
$K(q,q')=K(\widehat{q},\widehat{q}')$ on the BS kernel, where
$\widehat{q}_\mu= q_\mu- \frac{q.P}{P^2}P_\mu$ is the component of
internal momentum of the hadron that is orthogonal to the total
hadron momentum, i.e. $\widehat{q}.P=0$., while $\sigma
P_\mu=\frac{q.P}{P^2}P_\mu$ is the component of $q$ longitudinal
to $P$, where the 4-dimensional volume element is,
$d^4q=d^3\widehat{q}Md\sigma$, and following a sequence of steps
outlined in \cite{hluf16}, we get the covariant forms of four
Salpeter equations (in 4D variable $\widehat{q}$), which are
effective 3D forms of BSE, and are valid for hadrons in arbitrary
motion. The four independent Salpeter equations are \cite{hluf16}:
\begin{eqnarray}
 &&\nonumber(M-\omega_1-\omega_2)\psi^{++}(\hat{q})=\Lambda_{1}^{+}(\hat{q})\Gamma(\hat{q})\Lambda_{2}^{+}(\hat{q})\\&&
   \nonumber(M+\omega_1+\omega_2)\psi^{--}(\hat{q})=-\Lambda_{1}^{-}(\hat{q})\Gamma(\hat{q})\Lambda_{2}^{-}(\hat{q})\\&&
\nonumber \psi^{+-}(\hat{q})=0\\&&
 \psi^{-+}(\hat{q})=0\label{fw5}
\end{eqnarray}
where $\Lambda^{\pm}$  are the projection operators \cite{hluf16}
for each of the constituents. $\Gamma(\widehat{q})$ is the 4D
hadron-quark vertex function, which enters into the 4D BS wave
function, $\Psi(P,q) = S_{F}(p_{1})\Gamma(\hat{q})S_{F}(-p_{2})$,
where
\begin{equation}\label{6bb}
\Gamma(\hat{q})=\int \frac{d^{3}\hat{q}'}{(2\pi)^{3}}
K(\hat{q},\hat{q}')\psi(\hat{q}')
\end{equation}

We wish to emphasize that the present model after 3D reduction is
still covariant. This is due to the fact that we have reduced a
fully 4D BSE to 3D BSE (which are actually four Salpeter
equations) by use of Covariant Instantaneous Ansatz (CIA), which
is a Lorentz-invariant generalization of the Instantaneous
Approximation (IA). We thus obtain the covariant forms of Salpeter
equations, which are effective 3D forms of BSE, and are valid for
hadrons in arbitrary motion.

Regarding the interaction kernel $K(\hat q',\hat q)$
\cite{bhatnagar18}, it can be written as,
\begin{equation}\label{cf0}
 K(\hat q',\hat q)=(\frac{1}{2}\vec\lambda_1.\frac{1}{2}\vec\lambda_2)(\gamma_\mu\otimes\gamma_\mu)V(\hat q',\hat q)
\end{equation}
with colour, spin and orbital parts respectively. For a kernel
with the above spin dependence, we can rewrite the hadron-quark
vertex in Eq.(\ref{6bb}) as \cite{bhatnagar18},
\begin{equation}\label{6aa}
 \Gamma(\hat q)=\int\frac{d^3\hat q'}{(2\pi)^3}V(\hat q,\hat q')\gamma_\mu\psi(\hat q')\gamma_\mu,
\end{equation}
where, each of the $\gamma_{\mu}$s sandwich the BS wave function,
$\psi(\widehat{q})$, with the scalar part of the kernel,
$V=V_{OGE}+ V_{Confinement}$ as,

\begin{eqnarray}\label{fr1}
&&\nonumber V(\hat q,\hat q')=\frac{4\pi\alpha_s}{(\hat q-\hat
q')^2}
 +\frac{3}{4}\omega^2_{q\bar q}\int d^3r\bigg(\kappa r^2-\frac{C_0}{\omega_0^2}\bigg)e^{i(\hat q-\hat q').\vec r},\\&&
 \kappa=(1+4\hat m_1\hat m_2A_0M^2r^2)^{-\frac{1}{2}},
\end{eqnarray}
where, the confinement part with a sequence of steps can be
expressed as \\
$V_{c}(\hat q,\hat q')=-\frac{3}{4}(2\pi)^3\overline{ V}_{c}(\hat q)\delta^3(\hat q-\hat
q')$, with $\overline{ V}_{c}(\hat q)=\omega^2_{q\bar
q}\bigg(\kappa \vec \nabla^2_{\hat
q}+\frac{C_0}{\omega_0^2}\bigg)$, and $\kappa=(1-4\hat m_1\hat
m_2A_0M^2\vec\nabla^2_{\hat q})^{-\frac{1}{2}}$.\bigskip

The present work is a substantial improvement over our previous
works \cite{hluf16,bhatnagar18}, in the sense that we have taken
the full Dirac structure of the 3D BS wave function,
$\psi(\widehat{q})$ given in Eq.(\ref{uw1}) to calculate the spin part,
$\gamma_{\mu}\psi(\widehat{q})\gamma_{\mu}$ that enters into the
hadron-quark vertex function, $\Gamma(\hat{q})$ as well as the
right hand sides of the 3D coupled integral equations, in contrast
to \cite{hluf16,bhatnagar18}, where we took only the leading Dirac
structures in $\psi(\widehat{q})$ to evaluate
$\gamma_{\mu}\psi(\widehat{q})\gamma_{\mu}$, in the integrals on
the right of the Salpeter equations in \cite{hluf16,bhatnagar18}.
In this work, we have obtained the mass spectral equations with
this exact treatment of the spin part
$\gamma_{\mu}\Psi(\widehat{q})\gamma_{\mu}$. What we further find
is that the higher order terms of $\overline{V_c}$ that we had
ignored in \cite{hluf16, bhatnagar18} due to negligible
coefficients, $\omega_{q\overline{q}}^4$ associated with these
terms, get effectively cancelled out when we take the full Dirac
structure of the wave function, $\psi(\widehat{q})$.

Further, in the present work, we notice that with the use of the
exact treatment of the spin dependent part of the kernel in the
RHS of Salpeter equations, for case $m_1 = m_2$, we get the mass
spectrum of equal mass quarkonia ($\chi_{c0}, \eta_c$, and
$J/\Psi$), for both ground and excited states, where the excited
states are closer to data \cite{tanabashi2018} than the excited
states obtained in our previous works \cite{bhatnagar18,hluf16}.

The framework is quite general so far. Thus, to obtain the mass
spectral equation, we have to start with the above
four Salpeter equations to solve the instantaneous BS equation.\\

\section{Mass spectral equation for heavy-light scalar $0^{++}$ quarkonia}

We start with the general form of 4D BS wave function for scalar
meson ($0^{++}$) in \cite{smith69}. Then, making use of the 3D
reduction and making use of the fact that $\widehat{q}.P=0$, we
can write the general decomposition of the instantaneous BS wave
function for scalar mesons $(J^{pc}=0^{++})$, of dimensionality
$M$ being composed of various Dirac structures that are multiplied
with scalar functions $f_i(\hat q)$ and various powers of the
meson mass $M$ as \cite{bhatnagar18}
\begin{equation}\label{uw1}
 \psi^S(\hat q)=Mf_1(\hat q)-i{\not}Pf_2(\hat q)-i{\not}\hat qf_3(\hat q)-\frac{2{\not}P{\not}\hat q}{M}f_4(\hat q),
\end{equation}
Till now these amplitudes $f_{1}$, and $f_{4}$ in equation above
are all independent, and as per the power counting rule
\cite{bhatnagar06,bhatnagar14} proposed by us earlier, the $f_1$,
and $f_2$ are the amplitudes associated with the leading Dirac
structures, namely $M$ and ${\not}P$, while $f_3$ and $f_4$ will
be the amplitudes associated with the sub-leading Dirac
structures, namely, ${\not}\hat{q}$, and
$\frac{2{\not}P{\not}\hat{q}}{M}$.

We now use the last two Salpeter equations $\psi^{+-}(\hat q)=
\psi^{-+}(\hat q)=0$ in Eq.(\ref{fw5}), that can be used to obtain the
constraint relations between the scalar functions for unequal mass
mesons as
\begin{equation}
  f_1(\hat q)=\frac{-(m_1+m_2)\hat q^2}{M(\omega_1\omega_2+m_1m_2-\hat q^2)}f_3(\hat q),~~~~~~
  f_2(\hat q)=\frac{2(\omega_2-\omega_1)\hat q^2}{M(\omega_1m_2+m_1\omega_2)}f_4(\hat q)
\end{equation}
The BS-wave function for scalar mesons in Eq.(\ref{uw1}) with the
help of these constraint relations can be rewritten in terms of
only two independent scalar functions ($f_1$ (or $f_3$) and $f_4$)
as
\begin{equation}\label{uw11}
  \psi^S(\hat q)=\bigg(\frac{-(m_1+m_2)\hat q^2}{(\omega_1\omega_2+m_1m_2-\hat q^2)}-i{\not}\hat q\bigg)f_3(\hat q)\\
  -2\bigg(\frac{i(\omega_2-\omega_1)\hat q^2{\not}P}{M(\omega_1m_2+m_1\omega_2)}+\frac{{\not}P{\not}\hat q}{M}\bigg)f_4(\hat q)
\end{equation}
We wish to mention that due to these two above equations, the
scalar functions $f_i(\widehat{q}) (i= 1,...,4)$ are no longer all
independent, but are tied together by these relations, due to
which the amplitudes get mixed up\cite{bhatnagar18}.

The first two Salpeter equations of Eq.(\ref{fw5}) lead to a set
of coupled integral equations with the full structure of the wave
function $\psi^S(\hat q)$ in (\ref{uw11}) being used to evaluate
$\gamma_\mu\psi^S(\hat q)\gamma_\mu$ on the right hand sides of
these equations. We proceed in the same way as,
\cite{bhatnagar18}, where on the right side of these equations, we
first work with the confining interaction, $V_c(\widehat{q})$. We
show that these equations can be decoupled, and reduced to
algebraic equations in an approximate harmonic oscillator basis,
and solve them analytically. These equations are much more
involved than the equal mass case \cite{bhatnagar18}. We then
incorporate the one-gluon-exchange (OGE) term perturbatively, and
obtain the complete spectrum. These coupled equations (with use of
confining interaction alone) are,

\begin{equation}\label{co1}
\begin{split}
[M-\omega_1-\omega_2]\bigg[\frac{2\omega_1\omega_2(m_1+m_2)}{\omega_1\omega_2+m_1m_2
 -\hat q^2}f_3(\hat q)+\frac{4\omega_1\omega_2(m_1+m_2)}{\omega_1m_2+m_1\omega_2}f_4(\hat q)\bigg]
 =-\frac{1}{\hat q^2}\int\frac{d^3\hat q'}{(2\pi)^3}V_c(\hat q,\hat q')\\ \bigg[\bigg(\frac{4(\omega_1\omega_2-m_1m_2+\hat q^2)(m_1+m_2)\hat q'^2}
 {\omega_1\omega_2+m_1m_2-\hat q'^2} +2(m_1+m_2)\hat q.\hat q'\bigg)f_3(\hat q')\\
 -\bigg(\frac{4(\omega_1m_2-m_1\omega_2)(\omega_2-\omega_1)\hat q'^2}{\omega_1m_2+m_1\omega_2}\bigg)f_4(\hat q')\bigg],\\
 [M+\omega_1+\omega_2]\bigg[\frac{-2\omega_1\omega_2(m_1+m_2)}{\omega_1\omega_2+m_1m_2
 -\hat q^2}f_3(\hat q)+\frac{4\omega_1\omega_2(m_1+m_2)}{\omega_1m_2+m_1\omega_2}f_4(\hat q)\bigg]
 =-\frac{1}{\hat q^2}\int\frac{d^3\hat q'}{(2\pi)^3}V_c(\hat q,\hat q')\\
 \bigg[\bigg(\frac{4(\omega_1\omega_2-m_1m_2+\hat q^2)(m_1+m_2)\hat q'^2}
 {\omega_1\omega_2+m_1m_2-\hat q'^2} +2(m_1+m_2)\hat q.\hat q'\bigg)f_3(\hat q')\\
 +4\bigg(\frac{(\omega_1m_2-m_1\omega_2)(\omega_2-\omega_1)\hat q'^2}{\omega_1m_2+m_1\omega_2}\bigg)f_4(\hat q')\bigg]
 \end{split}
\end{equation}

To decouple these equations, we follow the same procedure in
\cite{hluf16,bhatnagar18}, where we first add them. Then we
subtract the second equation from the first equation. For a kernel
that can be expressed as
$V_c(\widehat{q}-\widehat{q}')=\overline{V}_c(\widehat{q})\delta^{3}(\widehat{q}-\widehat{q}')$,
we get two algebraic equations which are still coupled. Then from
one of the two equations so obtained, we eliminate
$f_{3}(\hat{q})$ in terms of $f_{4}(\hat{q})$, and plug this
expression for $f_{3}(\hat{q})$ in the second equation of the
coupled set so obtained to get a decoupled equation in
$f_{4}(\hat{q})$. Similarly, we eliminate $f_{4}(\hat{q})$ from
the second equation of the set of coupled algebraic equations in
terms of $f_{3}(\hat{q})$, and plug it into the first equation to
get a decoupled equation entirely in $f_{3}(\hat{q})$, which
reduces to two identical decoupled equations, one entirely in
$f_{3}(\hat{q})$, and the other that is entirely in
$f_{4}(\hat{q})$ as:

\begin{equation}
\begin{split}
  \bigg[\frac{M^2}{4}-\frac{1}{4}(m_1+m_2)^2-\hat q^2\bigg]f_3(\hat q)&=-\frac{1}{2}(m_1+m_2)\overline{ V}_{c}(\hat q)f_3(\hat q)\\
 \bigg[\frac{M^2}{4}-\frac{1}{4}(m_1+m_2)^2-\hat q^2\bigg]f_4(\hat q)&=-\frac{1}{2}(m_1+m_2)\overline{ V}_{c}(\hat q)f_4(\hat q)
\end{split}
\end{equation}

It is to be seen here that on RHS of the above two equations, with
the exact treatment of the spin part of the kernel, we get only
the terms that are linear in $\overline{V}_c$, (unlike
\cite{hluf16,bhatnagar18}, where we also obtained quadratic terms
of the type, $\overline{V}_c^2$, that were very small in magnitude
in comparison to $\overline{V_c}$). Since the two equations are of
the same form in scalar functions $f_3(\hat q)$ and $f_4(\hat q)$,
that are the solutions of identical equations, we can take,
$f_3(\hat q)\approx f_4(\hat q)(=\phi_S(\hat q))$. Using the
expression for $\overline{V}_c(\hat q)$ given above, we get the
equation,

\begin{equation}\label{s23}
 E_S\phi_S(\hat q)=[-\beta_S^4 \vec \nabla^2_{\hat q} +\hat q^2]\phi_S(\hat q),
\end{equation}

where the inverse range parameter $\beta_S$ can be expressed as,

\begin{equation}\label{be}
\begin{split}
  \beta_S &=\bigg(\frac{\frac{1}{2}\omega^2_{q\bar q}(m_1+m_2)}{\sqrt{1+8\hat m_1\hat m_2A_0(N+\frac{3}{2})}}\bigg)^{1/4},\\
 \omega_{q\bar q}&=(4M\hat m_1\hat m_2\omega_0^2\alpha_s(M))^{1/2},\\
  \alpha_s&=\frac{12\pi}{33-2N_f}\log\bigg(\frac{M^2}{\Lambda_{QCD}^2}\bigg)^{-1}
\end{split}
\end{equation}

Using the method of power series, this leads to the mass spectral
equation for scalar mesons as,

\begin{equation}\label{mse0}
\frac{1}{4}\bigg[M^2-(m_1+m_2)^2\bigg]+\frac{C_0\beta_S^4}{\omega_0^2}\sqrt{1+8\hat
m_1\hat
m_2A_0(N+\frac{3}{2})}=2\beta_S^{2}(N+\frac{3}{2}),~~~N=1,3,5,...,
\end{equation}

with the energy eigen value of the scalar mesons,
$E_S=2\beta_S^2(N+\frac{3}{2})$, where $N=2n+l$, with the
principal quantum number taking values $n=0,1,2,...$, and the
orbital quantum number $l=1$ that corresponds to $P$ wave states,
and the solutions of Eq.(\ref{s23}) are given by the following
normalized wave functions that are similar to the wave functions
in \cite{bhatnagar18}, except for the inverse range parameter
$\beta$ expression that is different from \cite{bhatnagar18} due
to the exact treatment of the spin part of the kernel, and also
the unequal mass kinematics. The overall structure of these wave
functions is very similar to the wave functions derived in
\cite{bhatnagar18}, except for the algebraic form of $\beta_S$.
They are:

\begin{equation}\label{wv1}
\begin{split}
 \phi_S(1P,\hat q)&=\sqrt{\frac{2}{3}}\frac{1}{\pi^{3/4}}\frac{1}{\beta_S^{5/2}} \hat qe^{-\frac{\hat q^2}{2\beta_S^2}}\\
 \phi_S(2P,\hat q)&=\sqrt{\frac{5}{3}}\frac{1}{\pi^{3/4}}\frac{1}{\beta_S^{5/2}}
  \hat q\bigg(1-\frac{2\hat q^2}{5\beta_S^2}\bigg)e^{-\frac{\hat q^2}{2\beta_S^2}}\\
    \phi_S(3P,\hat q)&=\sqrt{\frac{35}{12}}\frac{1}{\pi^{3/4}}\frac{1}{\beta_S^{5/2}}
 \hat q\bigg(1-\frac{4\hat q^2}{5\beta_S^2}+\frac{4\hat q^4}{35\beta_S^4}\bigg)e^{-\frac{\hat q^2}{2\beta_S^2}}\\
   \phi_S(4P,\hat q)&=\sqrt{\frac{35}{8}}\frac{1}{\pi^{3/4}}\frac{1}{\beta_S^{5/2}}
 \hat q\bigg(1-\frac{6\hat q^2}{5\beta_S^2}+\frac{12\hat q^4}{35\beta_S^4}-\frac{8\hat q^6}{315\beta_S^6}\bigg)e^{-\frac{\hat q^2}{2\beta_S^2}},
\end{split}
\end{equation}

Now, we  treat the mass spectral equation in Eq.(\ref{s23}), which
is obtained by taking only the confinement part of the kernel, as
an unperturbed spectral equation with the unperturbed wave
functions in Eq.(\ref{wv1}). We then incorporate  the one gluon
exchange term in the interaction kernel perturbatively (as in
\cite{bhatnagar18}) and solve to first order in perturbation
theory. The complete mass spectra of ground and excited states of
heavy-light scalar mesons is

\begin{equation}\label{mse0}
\frac{1}{8\beta_S^2}\bigg[M^2-(m_1+m_2)^2\bigg]+\frac{C_0\beta_S^2}{2\omega_0^2}\sqrt{1+8\hat
m_1\hat m_2A_0(N+\frac{3}{2})} +\gamma\langle V_{coul}^S\rangle
=N+\frac{3}{2},~~~N=1,3, 5,...,
\end{equation}

where $\langle V_{coul}^S\rangle$ is the expectation value of
$V_{coul}^S$ between the unperturbed states of the scalar mesons
with $l=1$ and $n=0,1,2,...$, and
$\gamma=\displaystyle\frac{\omega_0^4}{C_0\beta_S^2}(3l+1)$ is
introduced as a weighting factor to have the Coulomb term
dimensionally consistent with the harmonic term, and also acts as
a measure of the strength of the perturbation. The expectation
value of the Coulomb term associated with the OGE term for scalar
quarkonia for $nP  (n=1,2,3,...)$ states are

\begin{equation}
 \langle nP\mid V^S_{coul}\mid nP\rangle =-\frac{32\pi
 \alpha_s}{9\beta_S^2}.
\end{equation}

The results of our model for mass spectrum for scalar
$Q\overline{q}$ states along with data \cite{tanabashi2018}, and
other models is given in Table 1. It is observed that the mass
spectra of mesons of various $J^{PC}$ ($0^{++}, 0^{-+}$, and
$1^{--}$) is somewhat insensitive to a small range of variations
of parameter $\omega_0$, as long as $\frac{C_0}{\omega_0^2}$ is a
constant. The input parameters of our model obtained by best fit
to the spectra of ground states of scalar, pseudoscalar and vector
$Q\overline{q}$, and $Q\overline{Q}$ quarkonia are: $C_0$= 0.139,
$\omega_0$= 0.125 GeV, $\Lambda_{QCD}$= 0.200 GeV, and $A_0$=
0.01, with input quark masses $m_u$= 0.180 GeV, $m_s$= 0.350 GeV,
$m_c$= 1.490 GeV, and $m_b$= 5.070 GeV. Using these set of input
parameters, we do the mass spectral calculations of both ground
and excited states of heavy-light scalar ($0^{++}$) ( in section
3), pseudoscalar ($0^{-+}$) (in section 4) and vector
($1^{--}$)(in section 5) quarkonia.

\bigskip
 \begin{table}[htbp]
\begin{center}
\begin{tabular}{p{1.5cm} p{2cm} p{3.5cm} p{2cm} p{2cm} p{2.3cm} p{2cm}  }
\hline\hline
     &\footnotesize{BSE-CIA} &\small Expt.\cite{tanabashi2018}&\small BSE&\small~~ PM &\small Lattice QCD &\small~~ RQM  \\
\hline
\small $M_{B_c(1P_0)}$ &6.6722 & & &6.715\cite{Ajay06}&6.727$\pm $30\cite{davies96} &6.699\cite{ebert033} \\
\small $M_{B_c(2P_0)}$&7.2107& & &7.102\cite{Ajay06}  & &7.091\cite{ebert033}  \\
\small $M_{B_c(3P_0)}$&7.6030& & &  & &  \\
\small $M_{B_c(4P_0)}$&7.9286& & & & &   \\
\small $M_{B_c(5P_0)}$&8.2139 & & & & &  \\
\small $M_{B_c(6P_0)}$&8.4716 & & & & &   \\
\small $M_{B_s(1P_0)}$ &5.7057 & & &  5.812\cite{Virendrasinh} & & 5.833\cite{ebert10}   \\
\small $M_{B_s(2P_0)}$&6.0826 &  &  &  6.367\cite{Virendrasinh} &   &6.318\cite{ebert10}  \\
\small $M_{B_s(3P_0)}$&6.3707& &  &6.879\cite{Virendrasinh} & & \\
\small $M_{B_s(4P_0)}$&6.6154 & & & &  & \\
\small $M_{B_s(5P_0)}$&6.8329 & &  & & & \\
\small $M_{B_s(6P_0)}$&7.0309  & & & &  & \\
\small $M_{B(1P_0)}$ &5.5531 &  &  & 5.730\cite{Virendrasinh}  &   & 5.749\cite{ebert10}  \\
\small $M_{B(2P_0)}$&5.8972 &   &  & 6.297\cite{Virendrasinh}  &   &6.221\cite{ebert10}  \\
 \small $M_{B(3P_0)}$&6.1631& &  & 6.826\cite{Virendrasinh}& &   \\
 \small $M_{B(4P_0)}$&6.3901& & & &  & \\
 \small $M_{B(5P_0)}$&6.5924 & &  & & & \\
\small $M_{B(6P_0)}$&6.7770  & & & &  & \\
\small $M_{D(1P_0)}$ & 2.5088 &2.318$\pm$0.029  & &2.3864\cite{Vinodkumar}  &  &2.406\cite{ebert10}  \\
\small $M_{D(2P_0)}$&2.9200 & &  &2.8884\cite{Vinodkumar}  &    &2.919 \cite{ebert10} \\
\small $M_{D(3P_0)}$&3.2225&& & &        &  \\
\small $M_{D(4P_0)}$&3.4711& & & &      &   \\
\small $M_{D(5P_0)}$&3.6862 & &  & & & \\
\small $M_{D(6P_0)}$&3.8778  & & & &  & \\
\small $M_{D_s(1P_0)}$&  2.6229&2.3177$\pm$0.0006 &  &2.4945\cite{Vinodkumar}  & &2.509\cite{ebert10}\\
\small $M_{D_s(2P_0)}$&  3.0549 &  &   &3.0004\cite{Vinodkumar} & &3.054\cite{ebert10}\\
\small $M_{D_s(3P_0)}$& 3.3718 & & &&  &\\
\small $M_{D_s(4P_0)}$& 3.6323 & & & &&\\
\small $M_{D_s(5P_0)}$&3.8576 & &  & & & \\
\small $M_{D_s(6P_0)}$&4.0583  & & & &  & \\
\end{tabular}
\end{center}
\end{table}
\begin{table}[h!]
\begin{center}
\begin{tabular}{p{1.5cm} p{2cm} p{3.5cm} p{2cm} p{2cm} p{2.3cm} p{2cm}  }
$M_{\chi_{c0}}$(\footnotesize{$1P_0$})&3.4230&3.4147$\pm$0.00030   & &3.440\cite{Godfrey85}  & &3.413\cite{ebert11}   \\
$M_{\chi_{c0}}$(\footnotesize{$2P_0$})&4.0321&3.918$\pm$ 0.0019  &3.8368 \cite{glwang} &3.920\cite{Godfrey85}& &3.870\cite{ebert11} \\
$M_{\chi_{c0}}$(\footnotesize{$3P_0$})&4.4277& &  & & &4.301\cite{ebert11} \\
$M_{\chi_{c0}}$(\footnotesize{$4P_0$})&4.7514& & &&&\\
\hline \hline
\end{tabular}
\end{center}
\caption{Masss spectra of ground and excited states of scalar
$0^{++}$ quarkonia (in GeV)}
\end{table}
\bigskip

We now derive the mass spectral equations of unequal mass
pseudoscalar mesons in the next section.

\section{Mass spectral equations for heavy-light pseudoscalar $0^{-+}$ quarkonia}
The general decomposition for the 3D wave function of pseudoscalar
mesons obtained from the general 4D form \cite{smith69} through 3D
reduction as in previous section can be written as\cite{bhatnagar18}

\begin{equation}\label{wf2}
 \psi^P(\hat q)=[M\phi_1(\hat q)-i{\not}P\phi_2(\hat q)+i{\not}\hat q\phi_3(\hat q)+\frac{{\not}P{\not}\hat q}{M}\phi_4(\hat q)]\gamma_5
\end{equation}

We use the last two Salpeter equations in Eq.(\ref{fw5}) to find the
constraints on the components of the wave function as

\begin{equation}\label{cs4}
 \phi_4(\hat q)=\frac{M(m_1+m_2)}{\omega_1\omega_2+m_1m_2-\hat q^2}\phi_2(\hat q),~~~~~~~~~~
  \phi_3(\hat q)=\frac{M(\omega_1-\omega_2)}{\omega_1m_2+m_1\omega_2}\phi_1(\hat q),
\end{equation}

Plugging Eq.(\ref{cs4}) into Eq.(\ref{wf2}), we rewrite the wave function for pseudoscalar mesons as

\begin{equation}\label{wf4}
 \psi^P(\hat q)=\bigg[\bigg(M+\frac{iM(\omega_1-\omega_2)}{\omega_1m_2+m_1\omega_2}{\not}\hat q\bigg)\phi_1(\hat q)
 +\bigg(-i{\not}P+\frac{(m_1+m_2)}{\omega_1\omega_2 +m_1m_2-\hat q^2}{\not}P{\not}\hat q\bigg)\phi_2(\hat
 q)\bigg]\gamma_5.
\end{equation}

We use the first two Salpeter equations of Eq.(\ref{fw5}) to obtain the
corresponding coupled integral equations of pseudoscalar mesons
(with use of confining interaction alone) as,

\begin{equation}\label{rh9}
\begin{split}
[M-\omega_1-\omega_2]\bigg[\bigg(\frac{(m_2\omega_2)\omega^2_1+(m_1\omega_1)\omega^2_2}{\omega_1m_2+m_1\omega_2}\bigg)\phi_1(\hat
q)
 +\bigg(\frac{(m_2\omega_2)\omega^2_1+(m_1\omega_1)\omega^2_2}{\omega_1\omega_2 +m_1m_2-\hat q^2}\bigg)\phi_2(\hat q)\bigg]\\
=\int\frac{d^3\hat q'}{(2\pi)^3}V_c(\hat q,\hat q')\bigg[
 \bigg(-2(\omega_1\omega_2+m_1m_2+\hat q^2)-\frac{(m_1-m_2)(\omega_1-\omega_2)}{\omega_1m_2+m_1\omega_2}\hat q.\hat q'\bigg)\phi_1(\hat q')\\
 +(\omega_1m_2+m_1\omega_2)\phi_2(\hat q')\bigg]\\
[M+\omega_1+\omega_2]\bigg[\bigg(\frac{(m_2\omega_2)\omega^2_1+(m_1\omega_1)\omega^2_2}{\omega_1m_2+m_1\omega_2}\bigg)\phi_1(\hat
q)
 -\bigg(\frac{(m_2\omega_2)\omega^2_1+(m_1\omega_1)\omega^2_2}{\omega_1\omega_2 +m_1m_2-\hat q^2}\bigg)\phi_2(\hat q)\bigg]\\
=-\int\frac{d^3\hat q'}{(2\pi)^3}V_c(\hat q,\hat q')\bigg[
 \bigg(-2(\omega_1\omega_2+m_1m_2+\hat q^2)-\frac{(m_1-m_2)(\omega_1-\omega_2)}{\omega_1m_2+m_1\omega_2}\hat q.\hat q'\bigg)\phi_1(\hat q')\\
 -(\omega_1m_2+m_1\omega_2)\phi_2(\hat q')\bigg]
  \end{split}
\end{equation}

Using the same procedure as in the case of scalar mesons, these
two equations can be decoupled, and reduced to two independent
algebraic equations as

\begin{eqnarray}\label{nf4}
&&\nonumber
  [\frac{M^2}{4}-\frac{1}{4}(m_1+m_2)^2-\hat q^2]\phi_1(\hat q)=-\frac{1}{2}(m_1+m_2)\overline{ V}_{c}(\hat q)\phi_1(\hat q)
\\&&
\ [\frac{M^2}{4}-\frac{1}{4}(m_1+m_2)^2-\hat q^2]\phi_2(\hat
q)=-\frac{1}{2}(m_1+m_2)\overline{ V}_{c}(\hat q)\phi_2(\hat q).
\end{eqnarray}

Here, we again see that the scalar functions $\phi_1(\hat q)$ and
$\phi_2(\hat q)$ satisfy identical equations, and can be taken as
$\phi_1(\hat q)\approx \phi_2(\hat q)(=\phi_P(\hat q))$. Using the
expression for $\overline{V}_c(\hat q)$ after Eq.(\ref{fr1}), we obtain
the mass spectral equation as,

\begin{equation}\label{s2}
 E_P\phi_P(\hat q)=[-\beta_P^4 \vec \nabla^2_{\hat q} +\hat q^2]\phi_P(\hat q),
\end{equation}

whose solutions give the unperturbed mass spectrum (due to
confining interactions alone),

 \begin{equation}
 \frac{1}{8}\bigg[M^2-(m_1+m_2)^2\bigg]+\frac{C_0\beta_P^4}{2\omega_0^2}\sqrt{1+8\hat m_1\hat
 m_2A_0(N+\frac{3}{2})}=(N+\frac{3}{2})\beta_P^2;~~ N=2n+l,
 \end{equation}
with the orbital quantum number $l=0$ that corresponds to the $S$ states, and $\beta_P=\bigg(\frac{\frac{1}{2}\omega^2_{q\bar q}(m_1+m_2)}{\sqrt{1+8\hat m_1\hat m_2A_0(N+\frac{3}{2})}}\bigg)^{1/4}$. This unperturbed mass spectral equation of
pseudoscalar meson is the same as the corresponding spectral
equation of scalar meson in Eq.(\ref{s23}), except that $\beta_S$
is replaced by $\beta_P$, and $\phi_S(\hat q)$ replaced by
$\phi_P(\hat q)$. The normalized unperturbed wave functions of
$1S,...,4S$ states of pseudoscalar meson with $l=0$ are

\begin{equation}\label{25}
\begin{split}
  \phi_P(1S,\hat q)&=\frac{1}{\pi^{3/4}}\frac{1}{\beta_P^{3/2}}e^{-\frac{\hat q^2}{2\beta_P^2}}\\
 \phi_P(2S,\hat q)&=\sqrt{\frac{3}{2}}\frac{1}{\pi^{3/4}}\frac{1}{\beta_P^{3/2}}
  \bigg(1-\frac{2\hat q^2}{3\beta_P^2}\bigg)e^{-\frac{\hat q^2}{2\beta_P^2}}\\
 \phi_P(3S,\hat q)&=\sqrt{\frac{15}{8}}\frac{1}{\pi^{3/4}}\frac{1}{\beta_P^{3/2}}
  \bigg(1-\frac{4\hat q^2}{3\beta_P^2}+\frac{4\hat q^4}{15\beta_P^4}\bigg)e^{-\frac{\hat q^2}{2\beta_P^2}}\\
  \phi_P(4S,\hat q)&=\sqrt{\frac{35}{16}}\frac{1}{\pi^{3/4}}\frac{1}{\beta_P^{3/2}}
  \bigg(1-\frac{2\hat q^2}{\beta_P^2}+\frac{4\hat q^4}{5\beta_P^4}-\frac{8\hat q^6}{105\beta_P^6}\bigg)e^{-\frac{\hat q^2}{2\beta_P^2}}
  \end{split}
\end{equation}

We again incorporate the Coulomb term $V^P_{coul}$ associated with
the one gluon exchange interaction perturbatively into the
original mass spectral equation of pseudoscalar mesons, giving us
the complete mass spectra of ground and excited states of
heavy-light pseudoscalar quarkonia with orbital quantum number
$l=0$ as

\begin{equation}\label{mse0}
\frac{1}{8\beta_P^2}\bigg[M^2-(m_1+m_2)^2\bigg]+\frac{C_0\beta_P^2}{2\omega_0^2}\sqrt{1+8\hat m_1\hat m_2A_0(N+\frac{3}{2})}
+\gamma\langle V^P_{coul}\rangle =N+\frac{3}{2},~~~N=0,2,4,...,
\end{equation}

where again the perturbation parameter $\gamma$ has the same form
as in the case of scalar mesons with $\beta_S^2$ replaced by
$\beta_P^2$ and
$\gamma=\displaystyle\frac{\omega_0^4}{C_0\beta_P^2}(3l+1)$,
while, the first order correction to the total energy of the
system $E_P$ is given by the expectation value of the Coulomb term
between the unperturbed states of pseudoscalar mesons
$\phi_P(nS,\hat q)$ as

\begin{equation}
 \langle nS\mid V^{P}_{coul}\mid nS\rangle =-\frac{32\pi
 \alpha_s}{3\beta_{P}^2}.
\end{equation}

The results of our model for pseudoscalar $Q\overline{q}$ mesons
along with data\cite{tanabashi2018} and other models is given in
Table 2.
\bigskip

\begin{table}[htbp]
\begin{center}
\begin{tabular}{p{1.4cm} p{1.4cm} p{3.2cm} p{2.9cm} p{1.8cm} p{3.2cm} p{1.6cm}  }
\hline \hline
     &\footnotesize{BSE-CIA} &\small Expt.\cite{tanabashi2018}&\small{QCD Sum Rule}&\small~~ PM &\small Lattice QCD &\small RQM  \\
\hline
\small$M_{B_c(1S_0)}$ &6.2417  &6.2749$\pm$0.0008  &6.253\cite{gershtein} &6.349\cite{Ajay06}& 6.280$\pm$30$\pm$190\cite{davies96}&6.270\cite{ebert033} \\
\small$M_{B_c(2S_0)}$&6.9650 &                     &6.863\cite{gershtein} &6.821\cite{Ajay06} & 6.960$\pm $80\cite{davies96}&6.835\cite{ebert033} \\
\small$M_{B_c(3S_0)}$&7.4125 &                     &  &7.175\cite{Ajay06} & &7.193\cite{ebert033}  \\
\small$M_{B_c(4S_0)}$&7.7664 & &  & &  &  \\
\small $M_{B_c(5S_0)}$&8.0696 & &  & & & \\
\small $M_{B_c(6S_0)}$&8.3399  & & & &  & \\
\small$M_{B_s(1S_0)}$ &5.4211 & 5.3668$\pm$0.00019  &5.488$\pm$0.076 \cite{Halil18} &5.367\cite{Virendrasinh}  &   &5.372\cite{ebert10}  \\
\small$M_{B_s(2S_0)}$&5.9012 &  &  & 6.003\cite{Virendrasinh}  &  &5.976\cite{ebert10}   \\
\small$M_{B_s(3S_0)}$&6.2233 &    &  & 6.556 \cite{Virendrasinh} &   & 6.467\cite{ebert10}  \\
\small$M_{B_s(4S_0)}$&6.4864 & &  & 7.071 \cite{Virendrasinh} &   &  \\
\small $M_{B_s(5S_0)}$&6.7160 & &  & 7.565\cite{Virendrasinh}& & \\
\small $M_{B_s(6S_0)}$&6.9228  & & & &  & \\
\small$M_{B(1S_0)}$ &5.2955 &5.279$\pm$0.00014  &5.259$\pm$0.109\cite{Halil18}  &5.287\cite{Virendrasinh}   &   &5.280\cite{ebert10}   \\
\small$M_{B(2S_0)}$&5.7287 & &  & 5.926\cite{Virendrasinh}  &   &5.890\cite{ebert10}   \\
\small$M_{B(3S_0)}$&6.0245&  &  & 6.492\cite{Virendrasinh}  &   & 6.379\cite{ebert10}  \\
\small$M_{B(4S_0)}$&6.2679&  &  &7.027 \cite{Virendrasinh}  &   & \\
\small $M_{B(5S_0)}$&6.4811 & &  &7.538\cite{Virendrasinh} & & \\
\small $M_{B(6S_0)}$&6.6737  & & & &  & \\
\small$M_{D(1S_0)}$ &2.1390 &1.86906$\pm$00005  &1.972$\pm$0.094\cite{Halil18}  &1.8696\cite{Vinodkumar}  &    &1.871\cite{ebert10} \\
\small$M_{D(2S_0)}$&2.6743 & &  & 2.5235\cite{Vinodkumar} &   &2.581\cite{ebert10} \\
\small$M_{D(3S_0)}$&3.0199 & &  &   &    &3.062\cite{ebert10} \\
%
\small$M_{D(4S_0)}$&3.2923 &  &  &   &   & \\
\small $M_{D(5S_0)}$&3.5229 & &  & & & \\
\small $M_{D(6S_0)}$&3.7258  & & & &  & \\
\small$M_{D_s(1S_0)}$&2.2447 & 1.9683$\pm$0.00007   & &1.9686\cite{Vinodkumar}   &  &1.969\cite{ebert10}\\
\small$M_{D_s(2S_0)}$&   2.8095 &  &  &2.6333\cite{Vinodkumar}   & &2.688\cite{ebert10}\\
\small$M_{D_s(3S_0)}$&  3.1719  & &   &  &&3.129\cite{ebert10}\\
\small$M_{D_s(4S_0)}$&  3.4571 &  &  &   &&\\
\small $M_{D_s(5S_0)}$&3.6986 & &  & & & \\
\small $M_{D_s(6S_0)}$&3.9111  & & & &  & \\
\end{tabular}
\end{center}
\end{table}
\begin{table}[h!]
\begin{center}
\begin{tabular}{p{1.4cm} p{1.4cm} p{3.2cm} p{2.9cm} p{1.8cm} p{3.2cm} p{1.6cm}  }
$M_{\eta_c}$(\footnotesize{$1S_0$})&3.0132 &2.9839$\pm$0.0005&3.11$\pm$0.52\cite{Veli12} &2.980\cite{bhagyesh11} &3.292\cite{Burch10}&2.981\cite{ebert11} \\
$M_{\eta_c}$(\footnotesize{$2S_0$})&3.6910 &3.6376 $\pm$ 0.0012& &3.600\cite{bhagyesh11}&4.240\cite{Burch10} &3.635\cite{ebert11} \\
$M_{\eta_c}$(\footnotesize{$3S_0$})&4.1966&  & & 4.060\cite{bhagyesh11}& &3.986\cite{ebert11} \\
$M_{\eta_c}$(\footnotesize{$4S_0$})&4.5380 &  & & 4.4554\cite{bhagyesh11}& &4.401\cite{ebert11} \\
\hline \hline
\end{tabular}
\end{center}
\caption{Masss spectra of ground and excited states of
pseudoscalar $0^{-+}$ quarkonia (in GeV).}
\end{table}

We now give the derivation of the mass spectral equations of
vector mesons in the next section.

\section{Mass spectral equations for heavy-light vector $1^{--}$ quarkonia}
We again start with the general 4D decomposition \cite{smith69}.
Using 3D decomposition, the wave function of vector mesons can be
written as \cite{bhatnagar18, hluf16}:

\begin{equation}\label{wf5}
 \psi^V(\hat q)=iM{\not}\varepsilon\chi_1(\hat q)+{\not}\varepsilon{\not}P\chi_2(\hat q)+[{\not}\varepsilon{\not}\hat q-\hat q.\varepsilon]\chi_3(\hat q)
 -i[{\not}P{\not}\varepsilon{\not}\hat q+\hat q.\varepsilon{\not}P]\frac{1}{M}\chi_4(\hat q)+(\hat q.\varepsilon)\chi_5(\hat q)-i
 \hat q.\varepsilon\frac{{\not}P}{M}\chi_6(\hat q)
\end{equation}

The constraint equations on the components of the wave functions
($\chi$ s) can be obtained using the last two Salpeter equations
of (\ref{fw5}) as

\begin{equation}\label{bg1}
 \chi_5(\hat q)=\frac{M(m_1+m_2)}{\omega_1\omega_2+m_1m_2-\hat q^2}\chi_1(\hat q),
 ~~~~~~\chi_4(\hat q)=-\frac{M(\omega_1+\omega_2)}{2(\omega_1m_2+m_1\omega_2)}\chi_2(\hat q)
\end{equation}

\begin{equation*}\label{co5}
 \chi_3(\hat q)=\chi_6(\hat q)=0
\end{equation*}

Substituting Eq.(\ref{bg1}) into Eq.(\ref{wf5}), the wave function for vector mesons can be rewritten as
 \begin{equation}\label{wf6}
 \psi^V(\hat q)=\bigg(iM{\not}\varepsilon+\hat q.\varepsilon\frac{M(m_1+m_2)}{\omega_1\omega_2+m_1m_2-\hat q^2}
 \bigg)\chi_1(\hat q)+\bigg({\not}\varepsilon{\not}P+
 \frac{i(\omega_1+\omega_2)}{2(\omega_1m_2+m_1\omega_2)}({\not}P{\not}\varepsilon{\not}\hat q+\hat q.\varepsilon{\not}P)\bigg)\chi_2(\hat q)
\end{equation}

Using the first two Salpeter equations, we obtain the coupled
integral equations of vector mesons (with confining interaction alone) as

\begin{equation}\label{cov1}
\begin{split}
[M-\omega_1-\omega_2]\hat q.\varepsilon
 \bigg[\frac{2\omega_1\omega_2(m_1+m_2)}{\omega_1\omega_2+m_1m_2-\hat q^2}\chi_1(\hat q)
 -\frac{2\omega_1m_2(\omega_1+\omega_2)}{\omega_1m_2+m_1\omega_2}\chi_2(\hat q)\bigg]
 =\int\frac{d^3\hat q'}{(2\pi)^3}V_c(\hat q,\hat q')\\
 \bigg[-\bigg(2\hat q.\varepsilon(m_1+m_2)+4\hat q'.\varepsilon(m_1+m_2)\frac{(\omega_1\omega_2-m_1m_2+\hat q^2)}{\omega_1\omega_2+m_1m_2-\hat q'^2}\bigg)
 \chi_1(\hat q')\\
 +\bigg(2\hat q'.\varepsilon(\omega_1+\omega_2)\frac{(\omega_1m_2-m_1\omega_2)}{\omega_1m_2+m_1\omega_2}\bigg)\chi_2(\hat q')\bigg]\\
 [M+\omega_1+\omega_2]\hat q.\varepsilon
 \bigg[\frac{2\omega_1\omega_2(m_1+m_2)}{\omega_1\omega_2+m_1m_2-\hat q^2}\chi_1(\hat q)
 +\frac{2\omega_1m_2(\omega_1+\omega_2)}{\omega_1m_2+m_1\omega_2}\chi_2(\hat q)\bigg]
 =-\int\frac{d^3\hat q'}{(2\pi)^3}V_c(\hat q,\hat q')\\
 \bigg[-\bigg(2\hat q.\varepsilon(m_1+m_2)+4\hat q'.\varepsilon(m_1+m_2)\frac{(\omega_1\omega_2-m_1m_2+\hat q^2)}{\omega_1\omega_2+m_1m_2-\hat q'^2}\bigg)
 \chi_1(\hat q')\\
 -\bigg(2\hat q'.\varepsilon(\omega_1+\omega_2)\frac{(\omega_1m_2-m_1\omega_2)}{\omega_1m_2+m_1\omega_2}\bigg)\chi_2(\hat q')\bigg]
 \end{split}
\end{equation}

Now, using the same procedures to decouple these equations as in
the case of scalar and pseudoscalar mesons, we obtain two
decoupled algebraic equations,

\begin{equation}\label{32}
\begin{split}
[\frac{M^2}{4}-\frac{1}{4}(m_1+m_2)^2-\hat q^2]\chi_1(\hat q)&=-\frac{1}{2}(m_1+m_2)\overline{ V}_{c}(\hat q)\chi_1(\hat q)\\
 [\frac{M^2}{4}-\frac{1}{4}(m_1+m_2)^2-\hat q^2]\chi_2(\hat q)&=-\frac{1}{2}(m_1+m_2)\overline{ V}_{c}(\hat q)\chi_2(\hat q)\\
\end{split}
\end{equation}

Here, we see that the scalar functions $\chi_1(\hat q)$
and $\chi_2(\hat q)$ satisfy identical equations, and can be taken
as $\chi_1(\hat q)\approx \chi_2(\hat q)=\phi_V(\hat q)$. We then
obtain a single differential equation, which is nothing but the
equation of a simple quantum mechanical $3D$-harmonic oscillator
with coefficients depending on the hadron mass $M$, and total
quantum number $N$. The wave function satisfies the 3D BSE:

\begin{equation}\label{s1}
 \bigg[\frac{M^2}{4}-\frac{1}{4}(m_1+m_2)^2+\frac{C_0\beta_V^4}{\kappa\omega_0^2}\bigg]\phi_V(\hat q)
 =[-\beta_V^4 \vec \nabla^2_{\hat q} +\hat q^2]\phi_V(\hat q),
\end{equation}

which can be rewritten as
\begin{equation}\label{s2}
 E_V\phi_V(\hat q)=[-\beta_V^4 \vec \nabla^2_{\hat q} +\hat q^2]\phi_V(\hat q),
\end{equation}
where $\beta_V=\bigg(\frac{\omega^2_{q\bar
q}(m_1+m_2)}{\sqrt{1+8\hat m_1\hat
m_2A_0(N+\frac{3}{2})}}\bigg)^{1/4}$ is the inverse range
parameter, and the total energy of the system is identified as

\begin{equation}\label{s8}
 E_V=\frac{1}{4}\bigg[M^2-(m_1+m_2)^2\bigg]+\frac{C_0\beta_V^4}{\omega_0^2}\sqrt{1+8\hat m_1\hat m_2A_0(N+\frac{3}{2})}
\end{equation}

This mass spectral equation of vector meson is the same as the corresponding equation of scalar meson in Eq.(\ref{s23}), except that $\beta_S$
is replaced by $\beta_V$, and $\phi_S(\hat q)$ replaced by $\phi_V(\hat q)$. Therefore, the normalized wave functions of 1S,...,3D states
of vector meson, with S and D states corresponding to $l=0$ and $l=2$ respectively, are

\begin{equation}\label{sff}
\begin{split}
  \phi_V(1S,\hat q)&=\frac{1}{\pi^{3/4}}\frac{1}{\beta_V^{3/2}}e^{-\frac{\hat q^2}{2\beta_V^2}}\\
 \phi_V(2S,\hat q)&=\sqrt{\frac{3}{2}}\frac{1}{\pi^{3/4}}\frac{1}{\beta_V^{3/2}}
  \bigg(1-\frac{2\hat q^2}{3\beta_V^2}\bigg)e^{-\frac{\hat q^2}{2\beta_V^2}}\\
  \phi_V(1D,\hat q)&=\sqrt{\frac{4}{15}}\frac{1}{\pi^{3/4}}\frac{1}{\beta_V^{7/2}}\hat q^2e^{-\frac{\hat q^2}{2\beta_V^2}}\\
\phi_V(3S,\hat q)&=\sqrt{\frac{15}{8}}\frac{1}{\pi^{3/4}}\frac{1}{\beta_V^{3/2}}
     \bigg(1-\frac{4\hat q^2}{3\beta_V^2}+\frac{4\hat q^4}{15\beta_V^4}\bigg)e^{-\frac{\hat q^2}{2\beta_V^2}}\\
    \phi_V(2D,\hat q)&=\sqrt{\frac{14}{15}}\frac{1}{\pi^{3/4}}\frac{1}{\beta_V^{7/2}}
 \hat q^2 \bigg(1-\frac{2\hat q^2}{7\beta_V^2}\bigg)e^{-\frac{\hat q^2}{2\beta_V^2}}\\
    \phi_V(4S,\hat q)&=\sqrt{\frac{35}{16}}\frac{1}{\pi^{3/4}}\frac{1}{\beta_V^{3/2}}
  \bigg(1-\frac{2\hat q^2}{\beta_V^2}+\frac{4\hat q^4}{5\beta_V^4}-\frac{8\hat q^6}{105\beta_V^6}\bigg)e^{-\frac{\hat q^2}{2\beta_V^2}}\\
  \phi_V(3D,\hat q)&=\sqrt{\frac{21}{10}}\frac{1}{\pi^{3/4}}\frac{1}{\beta_V^{7/2}}
 \hat q^2 \bigg(1-\frac{4\hat q^2}{7\beta_V^2}+\frac{4\hat q^4}{63\beta_V^4}\bigg)e^{-\frac{\hat q^2}{2\beta_V^2}}\\
 \end{split}
\end{equation}

Eqs.(\ref{s2}- \ref{s8}) would lead to degenerate masses for $S$
and $D$ states of $Q\overline{q}$, and $Q\overline{Q}$ systems. To
get $S - D$ mass splitting, we make use of degenerate perturbation
theory. The Coulomb term $V^V_{coul}$ associated with the one
gluon exchange interaction is perturbatively incorporated into the
mass spectral equation, Eq.(\ref{s2}) (that is treated as the
unperturbed equation) for vector mesons, as:

\begin{equation}
 E_V\phi_V(\hat q)=[-\beta_V^4 \vec \nabla^2_{\hat q} +\hat q^2+V^V_{coul}]\phi_V(\hat q)
\end{equation}

The complete mass spectral equation of heavy-light vector quarkonia can be put as

\begin{equation}\label{mse0}
\frac{1}{8\beta^2_V}\bigg[M^2-(m_1+m_2)^2\bigg]+\frac{C_0\beta^2_V}{2\omega_0^2}\sqrt{1+8\hat m_1\hat m_2A_0(N+\frac{3}{2})}
+\gamma\langle V_{coul}^V\rangle =N+\frac{3}{2},~~~N=0,2,4,...,
\end{equation}

where $\langle V_{coul}^V\rangle$ has been weighted by the
perturbation parameter
$\gamma=\displaystyle\frac{\omega_0^4}{C_0\beta_V^2}(3l+1)$ (as in
the case of scalar $0^{++}$ and pseudo scalar $0^{-+}$ quarkonia),
and is given by the expectation value of the Coulomb term with
respect to the unperturbed states of vector mesons, in Eq.(\ref{sff}). In
the secular equation, the only non-zero expectation values of
$\langle V_{coul}^V\rangle$ are the ones that connect states of
the same quantum numbers, $n$ and $l$. They are:

\begin{equation}
\begin{split}
  \langle nS\mid V^{V}_{coul}\mid nS\rangle& =-\frac{32\pi \alpha_s}{3\beta_{V}^2}\\
 \langle nD\mid V^V_{coul}\mid nD\rangle &=-\frac{32\pi
 \alpha_s}{15\beta_V^2}
\end{split}
\end{equation}

We now give the plots of these normalized wave functions Vs.
$\widehat{q}$ (in Gev.) for different states of heavy pseudoscalar
and vector mesons ( such as composite of $c\overline{u}$,
$c\overline{s}$, $c\overline{c}$, $c\overline{b}$ and
$b\overline{b}$) in Fig.1-2 and in Fig.3-4, respectively. It can
be seen from these plots that the wave functions corresponding to
$nS$ and $nD$ states have $n-1$ nodes.
\begin{figure}[h]
\centering
\includegraphics[width=8cm]{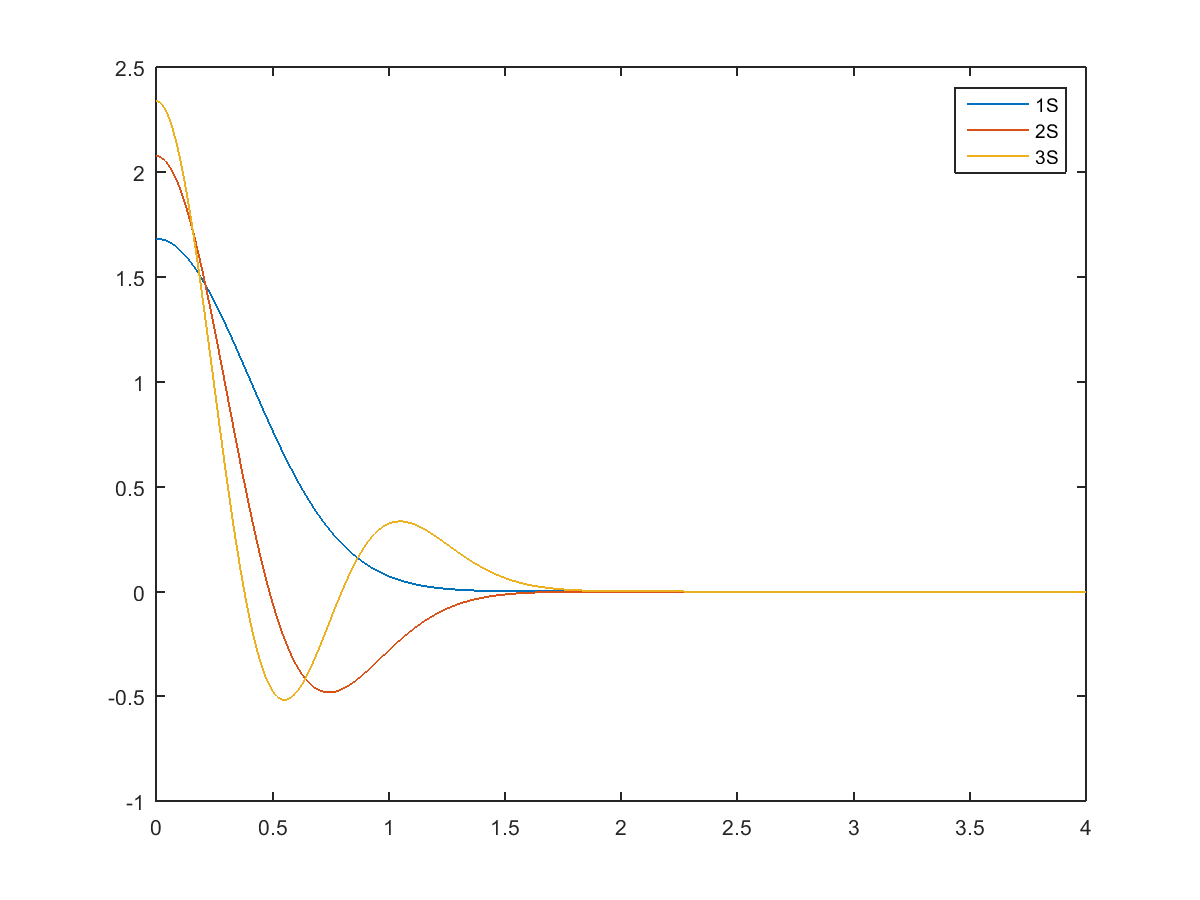}
\includegraphics[width=8cm]{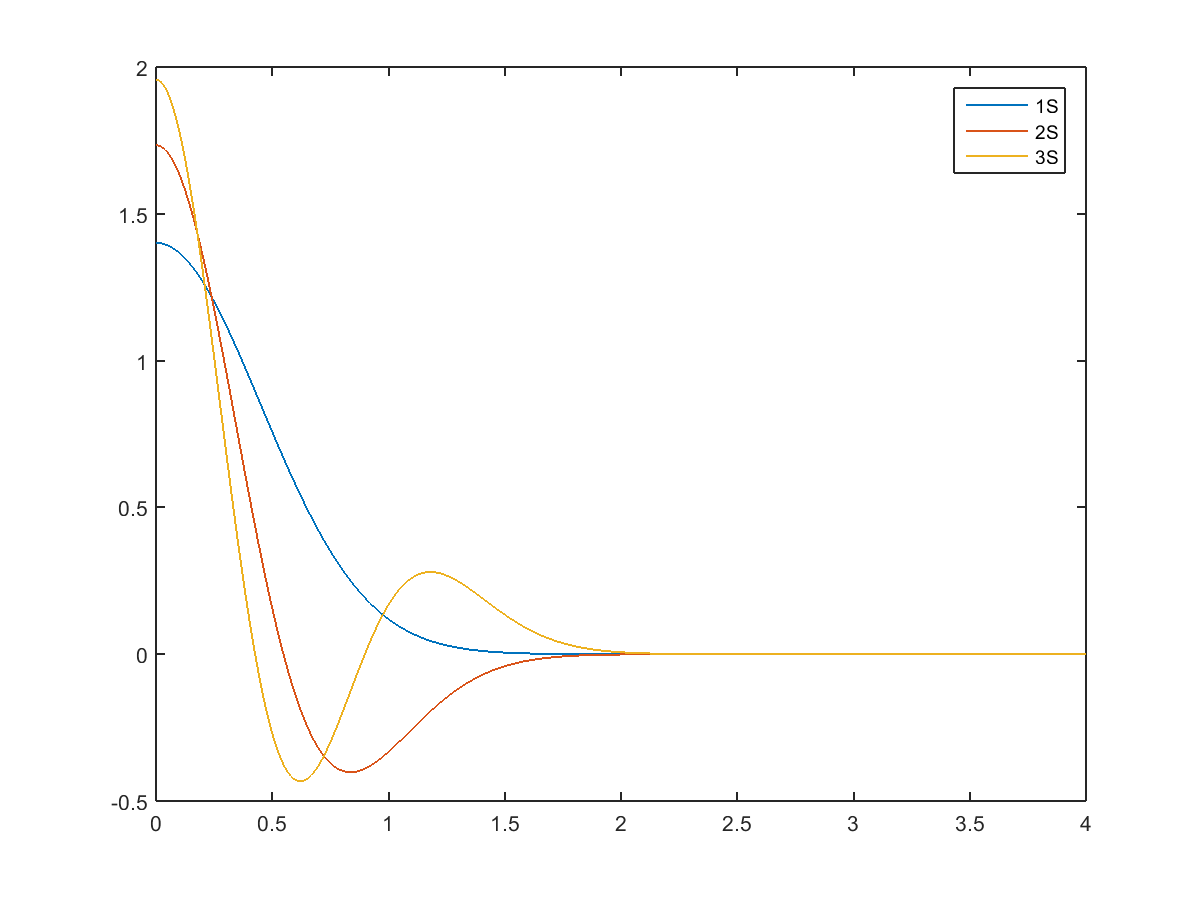}\\
\includegraphics[width=8cm]{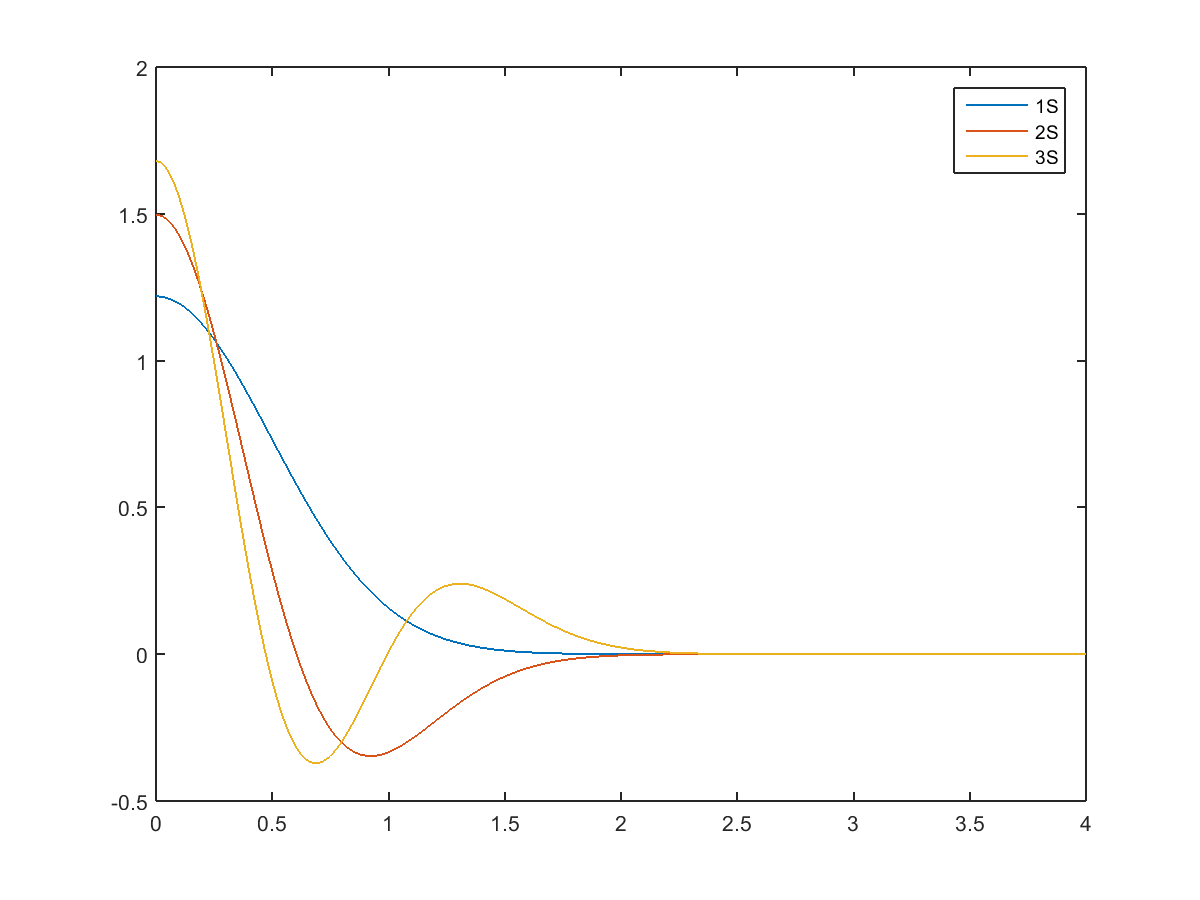}
\includegraphics[width=8cm]{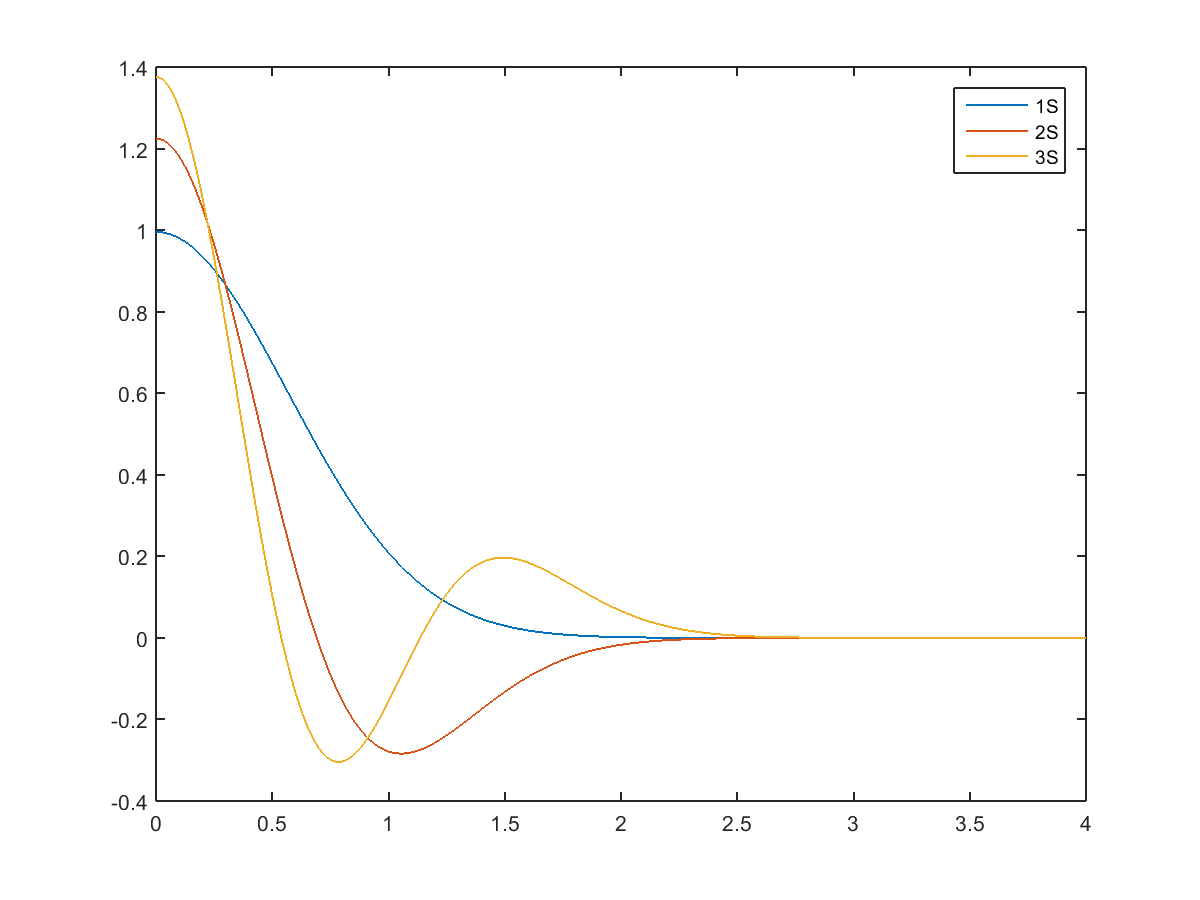}\\
\caption{Plots of wave functions for states $(1S,...,3S)$ Vs
$\widehat{q}$ (in Gev.) for pseudoscalar mesons, such as; $D$,
$D_{s}$, $B$ and $B_{s}$, respectively.}
\end{figure}

\begin{figure}[h]
\centering
\includegraphics[width=6cm]{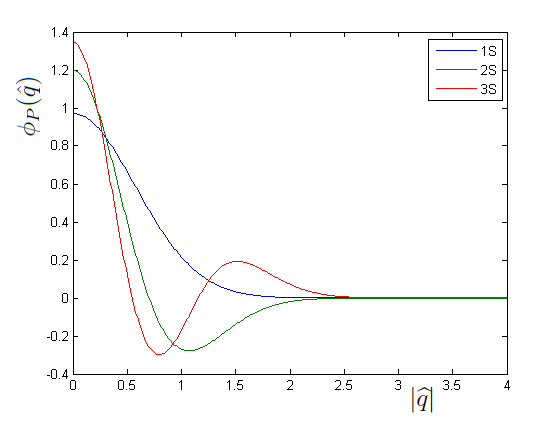}
\includegraphics[width=6cm]{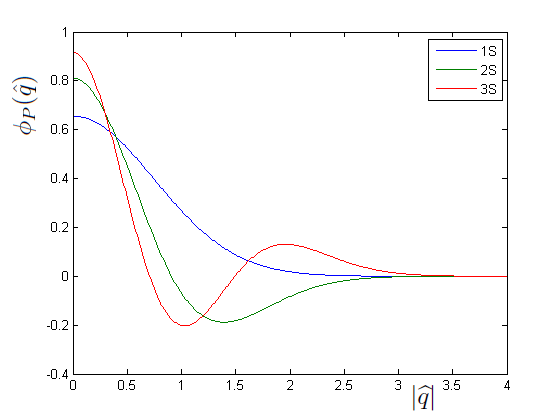}\\
\includegraphics[width=6cm]{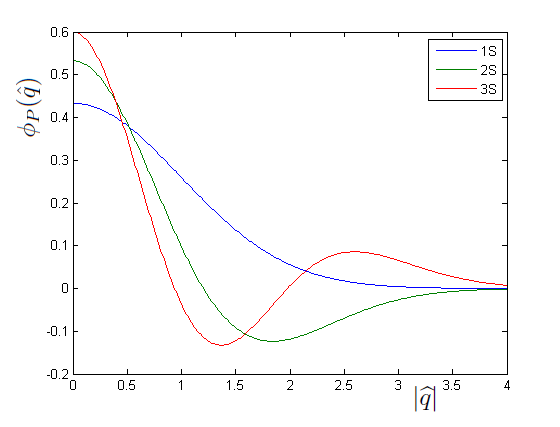}
\caption{Plots of wave functions for states $(1S,...,3S)$ Vs
$\widehat{q}$ (in Gev.) for pseudoscalar mesons, such as;
$\eta_{c}$, $B_{c}$ and $\eta_{b}$, respectively.}
\end{figure}
\begin{figure}[h]
\centering
\includegraphics[width=8cm]{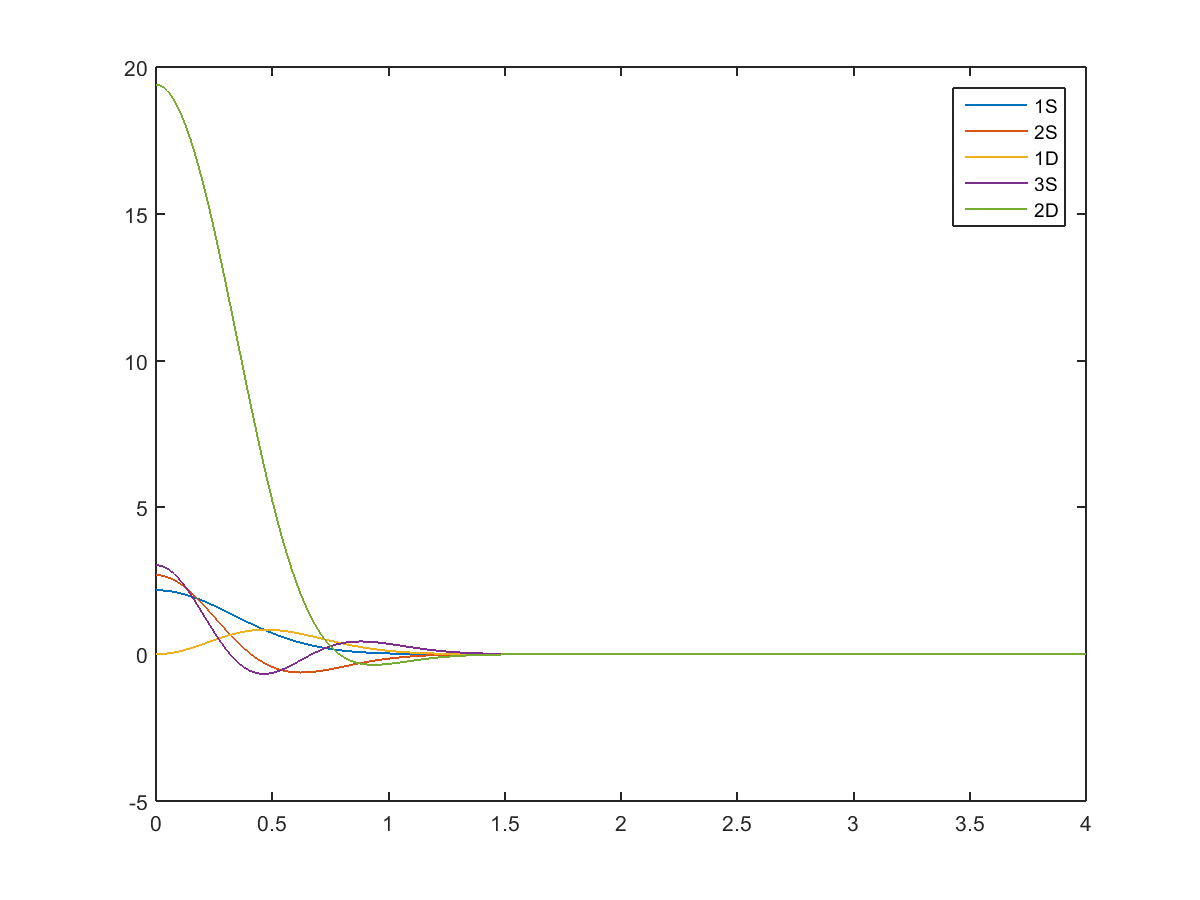}
\includegraphics[width=8cm]{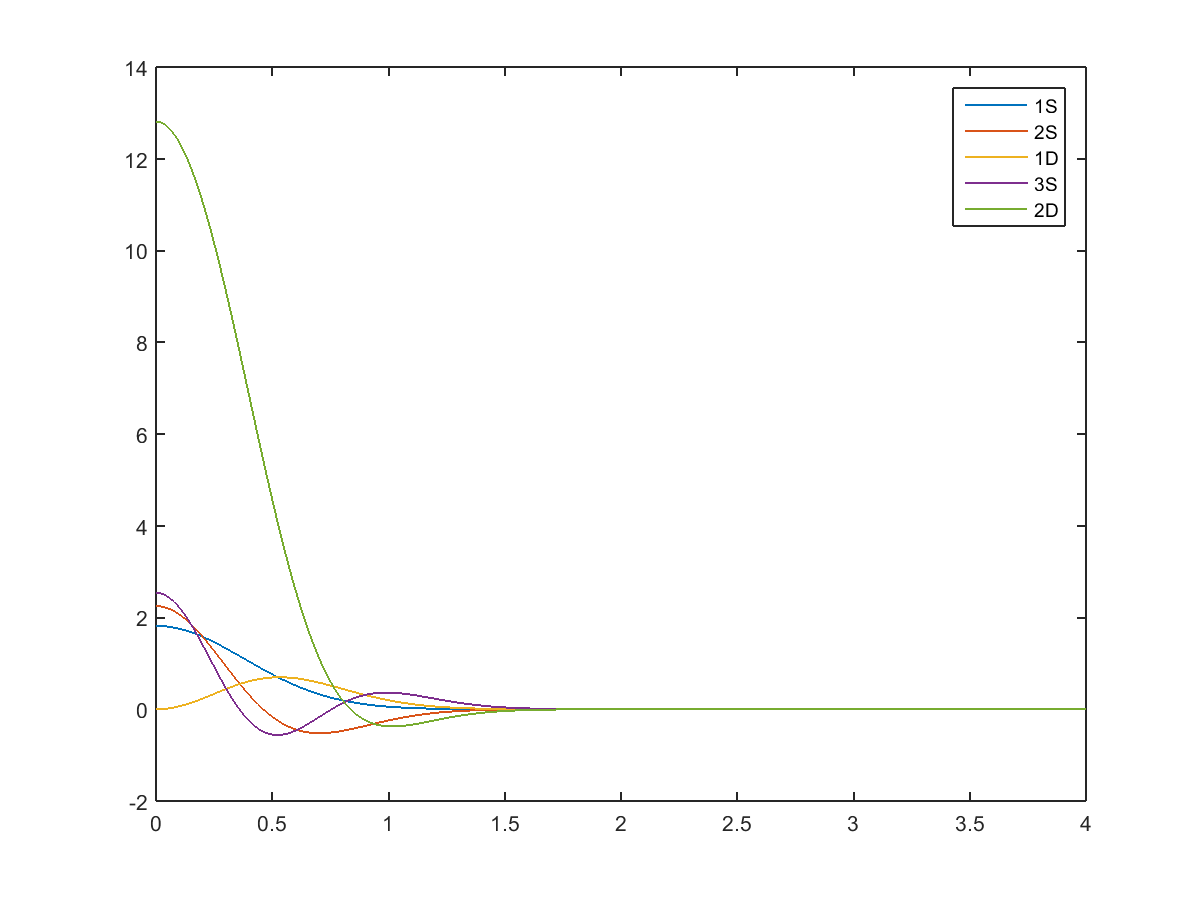}\\
\includegraphics[width=8cm]{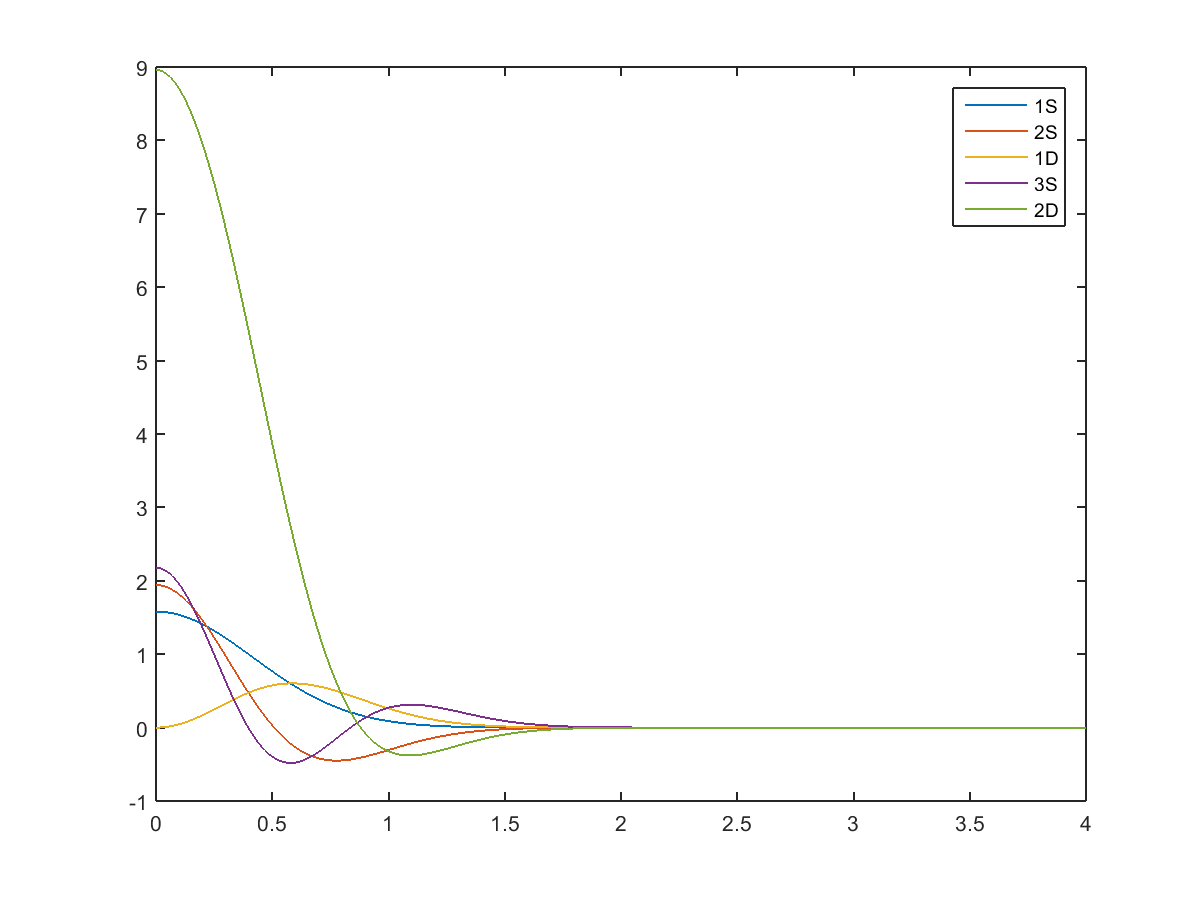}
\includegraphics[width=8cm]{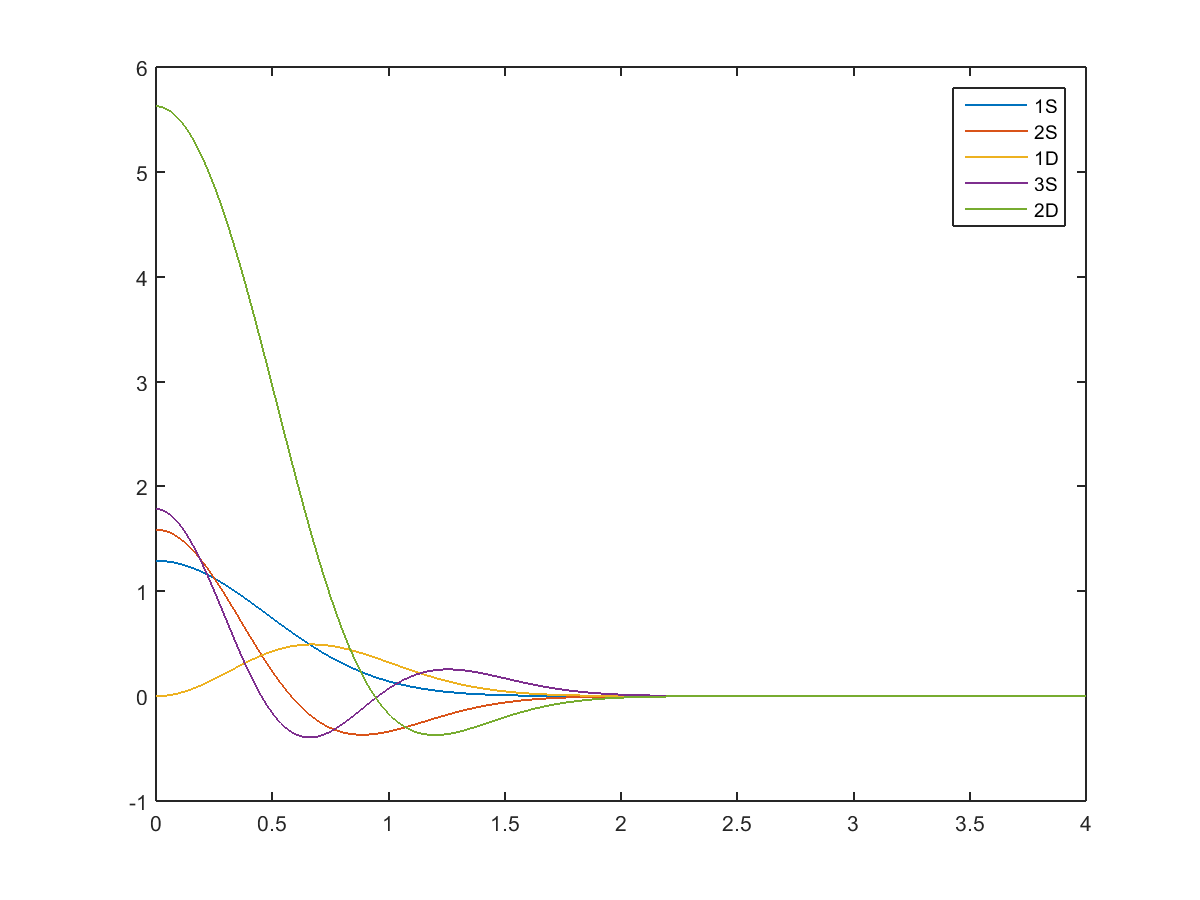}\\
\caption{Plots of wave functions for states $(1S,...,3S)$ Vs
$\widehat{q}$ (in Gev.) for vector mesons, such as; $D^{*}$,
$D^{*}_{s}$, $B^{*}$ and $B^{*}_{s}$, respectively.}
\end{figure}
\begin{figure}[h]
\centering
\includegraphics[width=6cm]{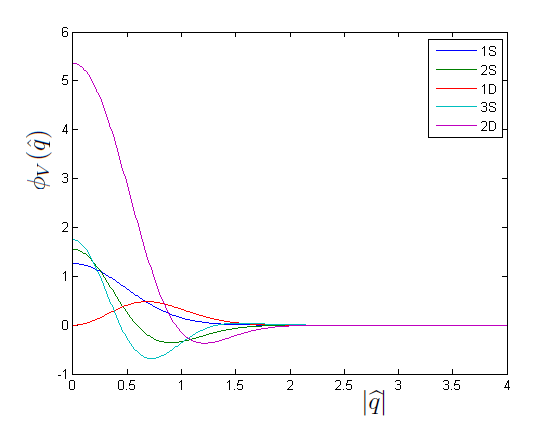}
\includegraphics[width=6cm]{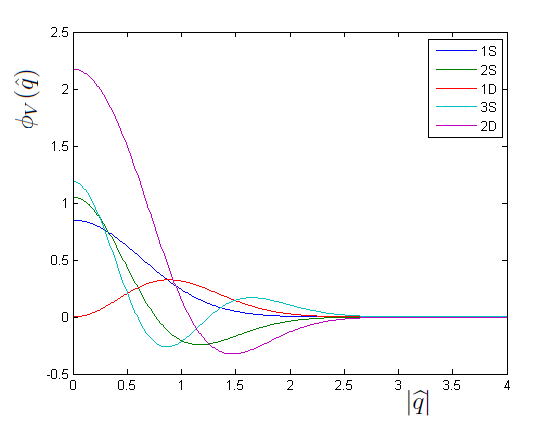}\\
\includegraphics[width=6cm]{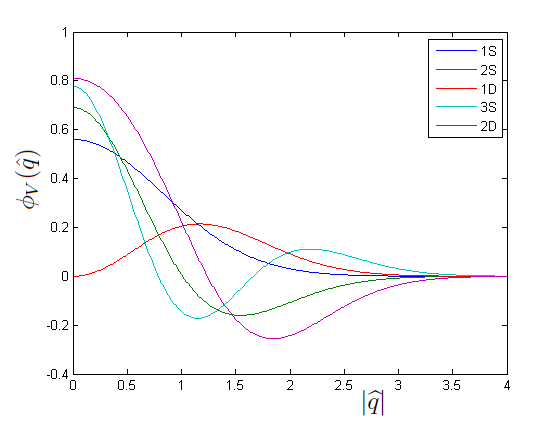}
\caption{Plots of wave functions for states $(1S,...,2D)$ Vs
$\widehat{q}$ (in Gev.) for vector mesons, such as; $J/\psi$,
$B^{*}_{c}$ and $\Upsilon$), respectively.}
\end{figure}

 \begin{table}[htbp]
\begin{center}
\begin{tabular}{p{1.5cm} p{2cm} p{3.7cm} p{2cm} p{2cm} p{2.7cm} p{2cm}  }
\hline \hline
      &\footnotesize{BSE-CIA} &\small Expt.\cite{tanabashi2018}&\small BSE&\small~~ PM &\small Lattice QCD&\small~~ RQM  \\
\hline
\small$M_{B_c^*(1S_1)}$&6.2417& & 6.3369\cite{Wang06}&6.373\cite{Ajay06} &6.321$\pm $20\cite{davies96} &6.332 \cite{ebert033}\\
\small$M_{B_c^*(2S_1)}$&6.9650 & &6.9185\cite{Wang06} &6.855\cite{Ajay06} &6.990$\pm $80\cite{davies96}  &6.881\cite{ebert033}\\
\small$M_{B_c^*(1D)}$&6.9712& & & & &7.072\cite{ebert033} \\
\small$M_{B_c^*(3S_1)}$&7.4125& & &7.210\cite{Ajay06} &  &7.235\cite{ebert033} \\
\small$M_{B_c^*(2D)}$&7.4190& & & &  &\\
\small$M_{B_c^*(4S_1)}$&7.7664& & & & &\\
%
%
\small$M_{B_c^*(3D)}$&7.7731& & &   & &\\
\small$M_{B_c^*(5S_1)}$&8.0696& & &  && \\
\small$M_{B_c^*(4D)}$&8.0765& & & && \\
\small$M_{B_c^*(6S_1)}$&8.3399& & &  && \\
\end{tabular}
\begin{tabular}{p{1.5cm} p{2cm} p{3.7cm} p{2cm} p{2cm} p{2.7cm} p{2cm}  }
\small$M_{B_s^*(1S_1)}$&5.4211&$5.4154^{+0.0014}_{-0.0015}$  &5.4166\cite{Wang06} &5.413\cite{Virendrasinh}  & &5.414\cite{ebert10} \\
\small$M_{B_s^*(2S_1)}$&5.9012& &5.9576\cite{Wang06} &6.029\cite{Virendrasinh} & &5.992\cite{ebert10}  \\
\small$M_{B_s^*(1D)}$&5.9129& & &6.119\cite{Virendrasinh}  & &6.209\cite{ebert10} \\
\small$M_{B_s^*(3S_1)}$&6.2233& & & 6.575\cite{Virendrasinh} & &6.475\cite{ebert10} \\
\small$M_{B_s^*(2D)}$&6.2357& &  &6.642\cite{Virendrasinh} & &6.629 \cite{ebert10} \\
\small$M_{B_s^*(4S_1)}$&6.4864& & &7.087\cite{Virendrasinh}& &\\
\small$M_{B_s^*(3D)}$&6.4993& &&7.139\cite{Virendrasinh}   & & \\
\small$M_{B_s^*(5S_1)}$&6.7160& &&7.579\cite{Virendrasinh} & & \\
\small$M_{B_s^*(4D)}$&6.7292& & &   & &  \\
\small$M_{B^*(1S_1)}$&5.4154&5.325$\pm$0.0004\cite{Olive14} &5.3229\cite{Wang06} &5.323\cite{Virendrasinh}   & &5.325\cite{ebert10}  \\
\small$M_{B^*(2S_1)}$&5.7287& &5.8377\cite{Wang06}& 5.947\cite{Virendrasinh}  & & 5.848\cite{ebert10}  \\
\small$M_{B^*(1D)}$&5.7422& &  &6.016\cite{Virendrasinh}  & &6.005\cite{ebert10}  \\
\small$M_{B^*(3S_1)}$&6.0245& & & 6.508\cite{Virendrasinh}& &6.136\cite{ebert10}  \\
\small$M_{B^*(2D)}$&6.0388& & & 6.562 \cite{Virendrasinh}  & &6.248\cite{ebert10}  \\
\small$M_{B^*(4S_1)}$&6.2679& & & 7.039\cite{Virendrasinh} & &\\
\small$M_{B^*(3D)}$&6.2828& & &  7.081\cite{Virendrasinh} & &  \\
\small$M_{B^*(5S_1)}$&6.4811& && 7.549\cite{Virendrasinh}& & \\
\small$M_{B^*(4D)}$&6.4963& & &   & &  \\
\small$M_{D^*(1S_1)}$&2.1390 &2.010$\pm$0.00005 &2.0065\cite{Wang06}  &2.0104\cite{Vinodkumar}  & & 2.010\cite{ebert10}   \\
\small$M_{D^*(2S_1)}$&2.6743&  &2.5408\cite{Wang06} & 2.6062\cite{Vinodkumar} &  &2.632 \cite{ebert10}     \\
\small$M_{D^*(1D)}$&2.7468& & &2.8029\cite{Vinodkumar} &  &    2.788\cite{ebert10}    \\
\small$M_{D^*(3S_1)}$&3.0199& & &3.1484\cite{Vinodkumar} & &  3.096\cite{ebert10}     \\
\end{tabular}
\end{center}
\end{table}

\begin{table}[h!]
\begin{center}
\begin{tabular}{p{1.5cm} p{2cm} p{3.7cm} p{2cm} p{2cm} p{2.7cm} p{2cm}  }
\small$M_{D^*(2D)}$&3.0919& & & 3.2818\cite{Vinodkumar} &   &  3.228\cite{ebert10}      \\
\small$M_{D^*(4S_1)}$&3.2923 & &  &   &   &     \\
\small$M_{D^*(3D)}$&3.3637& &  &   &      &      \\
\small$M_{D^*(5S_1)}$&3.5229 & &  & &     &    \\
\small$M_{D*(4D)}$&3.5938& &  & &     &       \\
\small$M_{D_s^*(1S_1)}$&  2.2447&2.1122$\pm$0.0004  &2.1120\cite{Wang06}  &2.1123\cite{Vinodkumar} & & 2.111\cite{ebert10}   \\
\small$M_{D_s^*(2S_1)}$&  2.8095&$2.7083^{+0.0040}_{-0.0034}$ &2.6730 \cite{Wang06} &2.7164\cite{Vinodkumar}  & & 2.731\cite{ebert10} \\
\small$M_{D_s^*(1D)}$&    2.8707& &   & 2.9145\cite{Vinodkumar} & & 2.919 \cite{ebert10}\\
\small$M_{D_s^*(3S_1)}$&   3.1719& &  &3.2626\cite{Vinodkumar}   & &3.242\cite{ebert10}   \\
\small$M_{D_s^*(2D)}$&    3.2326 & &  &3.3928\cite{Vinodkumar}  & &3.383\cite{ebert10}\\
\small$M_{D_s^*(4S_1)}$&   3.4571  &  &   & & &\\
\small$M_{D_s^*(3D)}$&    3.5174 &  & &  &  &\\
\small$M_{D_s^*(5S_1)}$&   3.6986& & && &\\
\small$M_{D_s^*(4D)}$&    3.7584 & & && &\\
$M_{J/\psi}$(\footnotesize{$1S_1$}) &3.0545&3.0969$\pm$ 0.000006  & &3.0969\cite{bhagyesh11}  & 3.099\cite{Kawanai16}&3.096\cite{ebert11}  \\
$M_{\psi}$(\footnotesize{$2S_1$})&3.7031&3.6861 $\pm$0.000025 &3.686\cite{glwang} &3.6890\cite{bhagyesh11}  & 3.653\cite{Kawanai16}&3.685\cite{ebert11}   \\
$M_{\psi}$(\footnotesize{1D})&3.7872&3.773$\pm$0.00033 &3.759\cite{glwang} & &  &3.783\cite{ebert11}  \\
$M_{\psi}$(\footnotesize{$3S_1$})&4.2037&4.039 $\pm$ 0.001 &4.065\cite{glwang}  &4.1407\cite{bhagyesh11} &4.099\cite{Kawanai16}&4.039\cite{ebert11}  \\
$M_{\psi}$(\footnotesize{2D})&4.2279&4.191 $\pm$0.005&4.108\cite{glwang} & & &4.150\cite{ebert11} \\
$M_{\psi}$(\footnotesize{$4S_1$})&4.5487&4.421 $\pm$ 0.004 &4.344\cite{glwang} &4.5320\cite{bhagyesh11}& &4.427\cite{ebert11} \\
$M_{\psi}$(\footnotesize{3D})&4.5729&  &4.371\cite{glwang}  & & &4.507\cite{ebert11} \\
$M_{\psi}$(\footnotesize{$5S_1$})&4.8408& &4.567\cite{glwang} &4.8841\cite{bhagyesh11}& &4.837\cite{ebert11} \\
$M_{\psi}$(\footnotesize{4D})&4.8649& & &  && 4.857\cite{ebert11} \\
\hline\hline
\end{tabular}
\end{center}
\caption{Masss spectra of ground and excited states of vector
$1^{--}$ quarkonia (in GeV)}
\end{table}

We now calculate the leptonic decays of pseudoscalar and vector
$Q\overline{q}$ mesons in the next sections.

\section{Leptonic decays of pseudoscalar heavy-light quarkonia}
The leptonic decays of pseudoscalar quarkonia proceed through the
coupling of the quark-anti quark loop to the axial vector current.
The leptonic decay constants, $f_P$ are defined as,

\begin{equation}
if_{P}P_{\mu} \equiv<0 |\bar{Q}i\gamma_{\mu}\gamma_{5}Q| P>.
\end{equation}

The decay constants can be expressed through the quark-loop
integral as,

\begin{equation}\label{41}
f_{P}P_{\mu}=\sqrt{3}\int \frac{d^{4}q}{(2\pi)^{4}}
Tr[\Psi^{P}(P,q)i\gamma_{\mu}\gamma_{5}].
\end{equation}

Here, the full 3D BS wave function, $\psi_P(\hat q)$ can be taken
from Eq.(\ref{wf4}), where $\phi_1(\hat q)$, and $\phi_2(\hat q)$ satisfy
two identical decoupled equations, Eq.(\ref{nf4}), leading to
$\phi_1(\hat q)=\phi_2(\hat q) (= \phi_P(\hat q))$, which is
expressed as,

\begin{equation}\label{wf44}
 \psi_P(\hat q)=N_P\bigg[M-i{\not}P+\frac{iM(\omega_1-\omega_2)}{\omega_1m_2+m_1\omega_2}{\not}\hat q
 +\frac{(m_1+m_2)}{\omega_1\omega_2 +m_1m_2-\hat q^2}{\not}P{\not}\hat q\bigg]\gamma_5\phi_P(\hat q)
\end{equation}

Putting the above expression for $\psi_P$ in Eq.(\ref{41}), and
evaluating trace over the gamma matrices on the right side of the
equation, we get,

\begin{equation}
f_{P}P_{\mu}=4\sqrt{3}\int
\frac{d^{4}q}{(2\pi)^{4}}\bigg[P_{\mu}-\frac{M(\omega_1-\omega_2)}{m_1\omega_1+m_2\omega_2}\widehat{q}_\mu\bigg]\phi_P(\hat
q).
\end{equation}

To evaluate $f_P$, we multiply both sides of the above equation by
$\frac{P_\mu}{M^2}$, and making use of the fact that
$\widehat{q}.P=0$, and the fact that,

\begin{equation}\label{3dr}
\phi_P(\hat q)=\int\frac{M d\sigma}{2\pi i}\Phi(p,q),
\end{equation}

we can express $f_P$ in terms of a 3D integral,

\begin{equation}
f_P=4\sqrt{3}N_P\int
\frac{d^{3}\widehat{q}}{(2\pi)^{3}}\phi_P(\hat q).
\end{equation}

where the 3D wave functions, $\phi_P(\hat q)$ for pseudoscalar
$Q\overline{q}$ states are given in Eqs.(\ref{25}), and $N_P$ is the 4D
BS normalizer obtained through the current conservation condition,

\begin{equation}\label{46}
2iP_\mu=\int \frac{d^{4}q}{(2\pi)^{4}}
\mbox{Tr}\left\{\overline{\Psi}(P,q)\left[\frac{\partial}{\partial
P_\mu}S_{F}^{-1}(p_1)\right]\Psi(P,q)S_{F}^{-1}(-p_2)\right\} +
(1\rightleftharpoons2),
\end{equation}

where $\Psi_P(\widehat{q})$ is the 4D BS wave function, while the
adjoint BS wave function, $\overline{\Psi}(P,q)=\gamma_4
\psi^{\dagger}(P,q)\gamma_4$. Making us of the fact that in the
inverse propagators, $S_F^{-1}(p_{1,2})$ of the two quarks, their
momenta are expressed as, $p_{1,2}=\widehat{m}_{1,2}P\pm q$, and
the 4D volume element, $d^4q=d^3\widehat{q}Md\sigma$. Integrating
over $Md\sigma$, and making us of the 3D form of BS wave function,
$\psi( \widehat{q})$, in Eq.(\ref{wf44}), evaluating trace over
the gamma matrices, and multiplying both sides of the equation by
$P_\mu$, we get,

\begin{equation}
\begin{split}
N_P^{-2}=\int \frac{d^3\widehat{q}}{(2\pi)^3}\phi_P^2(\widehat{q}
)[\frac{4M^2\widehat m_1\widehat
m_2(\omega_1-\omega_2)^2\widehat{q}^2}{(\omega_1 m_2+m_1
\omega_2)^2}+\frac{4M^2\widehat m_1\widehat m_2(m_1+m_2)^2\widehat{q}^2}{(\omega_1\omega_2+m_1m_2-\widehat{q}^2)^2}\\
+\frac{8M\widehat
m_1(m_1+m_2)m_2(\omega_1-\omega_2)\widehat{q}^2}{(\omega_1
m_2+m_1\omega_2)(\omega_1\omega_2+m_1m_2-\widehat{q}^2)}+\frac{8M\widehat
m_1(m_1+m_2)\widehat{q}^2
}{(\omega_1\omega_2+m_1m_2-\widehat{q}^2)}\\
+\frac{8M\widehat m_1(\omega_1-\omega_2)\widehat{q}^2
}{(\omega_1m_2+m_1\omega_2)}]+ (1\rightleftharpoons 2),
\end{split}
\end{equation}

where we take into account the contribution of the second term,
in Eq.(\ref{46}) to be the same as the contribution
of the first term. Leptonic decay constants of $0^{-+}$ quarkonia
are given in Table 4 along with data and results of other models.

\begin{table}[htbp]
\begin{center}
\begin{tabular}{p{0.8cm} p{1.6cm} p{3.2cm} p{1.6cm} p{3.2cm} p{3cm} p{2.3cm} }
  \hline\hline
   &BSE-CIA&Expt.&BSE\cite{wang044}& QCD SR&Latt. QCD &Rel. PM\cite{mao12}\\
   \hline
    $f_{\eta_{c}(1S)}$&0.4034& 0.335$\pm$0.075\cite{cleo01}& &0.260$\pm$0.075\cite{Veli12} &0.3928\cite{McNielle}&\\
    $f_{\eta_{c}(2S)}$&0.3068&  &  &  & & \\
    $f_{\eta_{c}(3S)}$&0.2660&  & &   & & \\

     $f_{B_c(1S)}$&0.3152 &  &  &0.400$\pm$0.015\cite{Kiselev} & &\\
    $f_{B_c(2S)}$&0.2459&  &  &   &  & \\
    $f_{B_c(3S)}$&0.2170 &  &  &  & & \\

     $f_{B_s(1S)}$&0.1917 &  &  &0.195\cite{wang044} & &0.2288$\pm$0.0069\\
    $f_{B_s(2S)}$&0.1610 &  &    & &  & \\
    $f_{B_s(3S)}$&0.1470&  &  &  &  & \\

     $f_{B(1S)}$&0.1691 &  & 0.192 &0.1915$\pm$0.0073\cite{lucha13} & &0.198$\pm$0.014\\
    $f_{B(2S)}$&0.1456&  &    &  &  & \\
    $f_{B(3S)}$&0.1342 &  &  &  &  &\\

     $f_{D_s(1S)}$&0.2428&0.2546$\pm$0.0059\cite{Eisenstein08} &  & &0.241$\pm$0.0003\cite{Follana08} &0.256$\pm$0.026\\
    $f_{D_s(2S)}$&0.1945&  & &   &   & \\
    $f_{D_s(3S)}$&0.1730&  &  &  &   &\\

    $f_{D(1S)}$&0.2088 &0.2067$\pm$0.0089\cite{oliveber1412}  & &  &0.207$\pm$0.0004\cite{Follana08} &0.208$\pm$0.021 \\
    $f_{D(2S)}$&0.1724 &  &  &   &   &\\
    $f_{D(3S)}$&0.1550&  &  & &  &  \\ \hline
   \hline
   \end{tabular}
   \end{center}
   \caption{Leptonic decay constants, $f_P$ of ground
state (1S) and excited state (2S) and (3S) of heavy-light
pseudoscalar mesons (in GeV.) in present calculation (BSE-CIA)
along with experimental data, and their masses in other models.}
     \end{table}

\section{Leptonic decays of heavy-light vector quarkonia}
Leptonic decays of vector quarkonia are defined through the
equation,

\begin{equation}
f_{V}M\epsilon_{\mu}(P) \equiv<0 |\bar{Q}\gamma_{\mu}Q| V(P)>
\end{equation}
The decay constant $f_V$ can be expressed through the quark loop
integral,
\begin{equation}\label{wf666}
f_{V}M\epsilon_{\mu} =\sqrt{3}\int \frac{d^{3}\hat{q}}{(2\pi)^{3}}
Tr[\psi^{V}(\hat{q})\gamma_{\mu}].
\end{equation}
Here $\psi^{V}(\hat{q})$ is the full 3D wave function for vector
mesons that can be taken from Eq.(\ref{wf6}), where $\chi_1(\hat q)$, and
$\chi_2(\hat q)$ satisfy two identical decoupled equations,
Eq.(\ref{32}), leading to $\chi_1(\hat q)=\chi_2(\hat q) (= \phi_V(\hat
q))$, which is expressed as,
\begin{equation}\label{wf66}
 \psi_V(\hat q)=N_V\bigg[iM{\not}\varepsilon+\hat q.\varepsilon\frac{M(m_1+m_2)}{\omega_1\omega_2+m_1m_2-\hat q^2}
 +{\not}\varepsilon{\not}P+
 \frac{i(\omega_1-\omega_2)}{2(\omega_1m_2+m_1\omega_2)}({\not}P{\not}\varepsilon{\not}\hat q+\hat q.\varepsilon{\not}P)\bigg]\phi_V(\hat q)
\end{equation}

Putting Eq.(\ref{wf66}) in Eq.(\ref{wf666}), and evaluating trace over the gamma
matrices on the RHS, and multiplying both sides of the resulting
equation by the polarization vector, $\epsilon_{\mu}$ of vector
meson, and making use of the fact that, $P.\epsilon=0$, and the 3D
reduction through Eq.(\ref{3dr}), we get the leptonic decay constant of
vector mesons as,

\begin{equation}
f_{V}=4\sqrt{3}N_{V}\int
\frac{d^{3}\hat{q}}{(2\pi)^{3}}\phi_{V}(\hat{q}),
\end{equation}

where the 4D BS normalizer, $N_V$ can be obtained from the current conservation condition in Eq.(\ref{46}), and following a similar procedure as in the case of pseudoscalar quarkonia as

\begin{equation}
\begin{split}
N_V^{-2}=\int
\frac{d^3\widehat{q}}{(2\pi)^3}\phi_V^2(\widehat{q})
[-\frac{2M^2\widehat{m_1}\widehat{m_2}(\omega_1+\omega_2)^2(\widehat{q}.\epsilon)^2}{(\omega_1 m_2+m_1
\omega_2)^2}-\frac{8M^2\widehat{m_1}\widehat{m_2}(m_1+m_2)^2(\widehat{q}.\epsilon)^2}{(\omega_1\omega_2+m_1m_2-\widehat{q}^2)^2}\\
+\frac{16M\widehat{m_1}(m_1+m_2)m_2(\omega_1+\omega_2)(\widehat{q}.\epsilon)^2}{(\omega_1
m_2+m_1\omega_2)(\omega_1\omega_2+m_1m_2-\widehat{q}^2)}+\frac{8M\widehat{m_1}(\omega_1+\omega_2)(\widehat{q}.\epsilon)^2 }{(\omega_1 m_2+m_1\omega_2)}\\
+\frac{16M\widehat{m_1}(m_1+m_2)(\widehat{q}.\epsilon)^2
}{(\omega_1\omega_2+m_1m_2-\widehat{q}^2)}]+(1 \rightleftharpoons
2),
\end{split}
\end{equation}
The leptonic decay constants of heavy-light vector mesons are
given in Table 5.

\begin{table}[htbp]
\begin{center}
\begin{tabular}{p{1.5cm} p{2cm} p{2.5cm} p{2.5cm} p{2cm}}
  \hline\hline
   &BSE - CIA &Expt.\cite{Olive14}&BSE\cite{Wang06}&RQM\cite{Ebert02}\\
   \hline
     $f_{Bc*(1S)}$&0.4679& &0.418$\pm$0.024 & \\
    $f_{Bc*(2S)}$&0.3622&  & 0.331$\pm$0.021 &    \\
    $f_{Bc*(1D)}$&0.4554&  &   &   \\
    $f_{Bc*(3S)}$ &0.3180 &  & & \\
    $f_{Bs*(1S)}$&0.2919& &0.272$\pm$0.020 &0.214 \\
    $f_{Bs*(2S)}$&0.2415&  &0.246$\pm$0.013  &    \\
    $f_{Bs*(1D)}$&0.2989&  &   &   \\
    $f_{Bs*(3S)}$ &0.2189  &  & & \\

    $f_{B*(1S)}$&0.2627& &0.238$\pm$0.018 &0.195\\
    $f_{B*(2S)}$&0.2220&  &0.221$\pm$0.014   &  \\
    $f_{B*(1D)}$&0.2737&  &    &  \\
    $f_{B*(3S)}$ &0.2030  &  & & \\

    $f_{Ds*(1S)}$&0.3892& &0.375$\pm$0.024 &0.335 \\
    $f_{Ds*(2S)}$&0.3060&  & 0.312$\pm$0.017  &   \\
    $f_{Ds*(1D)}$&0.3836&  &   &   \\
    $f_{Ds*(3S)}$ &0.2697  &  & & \\

     $f_{D*(1S)}$&0.3615& &0.339$\pm$0.022 &0.315\\
    $f_{D*(2S)}$&0.2888&  & 0.289$\pm$0.016   &  \\
    $f_{D*(1D)}$&0.3608&  &     & \\
    $f_{D*(3S)}$ &0.2562  &  & &\\

    $f_{J/\psi(1S)}$&0.5055& 0.411$\pm$0.007& & \\
    $f_{\psi(2S)}$&0.3300& 0.279$\pm$0.008 &   &   \\
    $f_{\psi(1D)}$&0.2439& 0.210$\pm$0.00024 &  &    \\
    $f_{\psi(3S)}$ &0.2128 &  & & \\
    \hline
   \hline
   \end{tabular}
   \end{center}
   \caption{Leptonic decay constants, $f_V$ of ground
state (1S) and excited state (2S),...,(3S) of heavy-light vector
mesons (in GeV.) in present calculation (BSE-CIA) along with
experimental data, and their masses in other models.}
     \end{table}

\section{Results and Discussion}
We have employed a 3D reduction of BSE (with a 4$\times$4
representation for two-body ($q\bar q$) BS amplitude) under
Covariant Instantaneous Ansatz (CIA) with an interaction kernel
consisting of both the confining and one gluon exchange terms, to
derive the algebraic forms of the mass spectral equations and
eigen functions of heavy-light quarkonia in an approximate
harmonic oscillator basis, leading to mass spectra of ground and
excited states of heavy-light scalar ($0^{++}$), pseudoscalar
($0^{-+}$), and vector ($1^{--}$) quarkonia.  And an interesting
feature of our analytic approach is that the plots of the
algebraic forms of our wave functions Eqs. (25) for pseudoscalar
quarkonia (in Figs.1-2), and Eqs.(36) for vector quarkonia (in
Figs. 3-4) respectively,  are very similar to the corresponding
plots of wave functions obtained in purely numerical approaches in
\cite{glwang}, and hence validating the correctness of our
analytical approach. These wave functions for heavy-light mesons
so derived, are then used to calculate the leptonic decay
constants for heavy-light pseudoscalar and vector mesons as a test
of the wave functions derived and the BSE framework employed.

As stated earlier, the partitioning of relativistic internal
momentum $q$ comes from the Wightmann-Garding definitions
$\widehat{m}_{1,2}$ of masses of individual quarks. The 3D
reduction through Covariant Instantaneous Ansatz (CIA) employed by
us does make our formulation relativistically covariant, but it
may not be Poincare covariant, since our results may depend on the
choice of internal momentum. This feature in our framework is yet
to be explored in detail, since in a Poincare covariant framework,
\cite{hilger15a, hilger15b}, the numerical results for the
amplitudes and masses are independent of the choice of momentum
partitioning parameters. This detailed study we intend to do next.

In this work, we make use of the exact treatment of the spin
structure $(\gamma_{\mu}\bigotimes\gamma_{\mu})$ in the
interaction kernel, in contrast to the approximate treatment of
the same in our previous works \cite{hluf16, bhatnagar18}). In so
doing we do away with the approximation of taking the leading
Dirac structures in the structure of 4D BS wave function,
$\Psi(P,q)$, which is a substantial improvement over our previous
works. We thus first derive analytically the mass spectral
equation using only the confining part of the interaction kernel
for $Q\overline{q}$ systems, and calculate the algebraic forms of
the wave functions. Then treating this mass spectral equation as
the unperturbed equation, we introduce the One-Gluon-Exchange
(OGE) perturbatively, and obtain the mass spectra for various
states of $0^{++}, 0^{-+}$, and $1^{--}$ , treating the wave
functions derived above as the unperturbed wave functions.

As mentioned earlier, in our works, we are not only interested in
studying the mass spectrum of hadrons, which no doubt is an
important element to study dynamics of hadrons, but also the
hadronic wave functions that play an important role in the
calculation of decay constants, form factors, structure functions
etc. for $Q\overline{Q}$, and $Q\overline{q}$ hadrons, and so far,
one of the central difficulties in tests of QCD is lack of
knowledge of hadronic wave functions. These hadronic
Bethe-Salpeter wave functions calculated algebraically in this
work can act as a bridge between the long distance
non-perturbative physics, and the short distance perturbative
physics. And since these quarkonia are involved in a number of
reactions which are of great importance for study of
Cabibbo-Kobayashi-Maskawa (CKM) matrix and CP violation, the wave
functions calculated analytically by us can lead to studies on a
number of processes involving $Q\overline{Q}$, and $Q\overline{q}$
states.  In this work, we have used these algebraic forms of wave
functions to calculate the leptonic decay constants of
pseudoscalar and vector $Q\overline{q}$ quarkonia in Tables 4 and
5 respectively.

We have first obtained the numerical values of masses for ground
and excited states of various heavy-light mesons and made
comparison of our results with experimental data and other models.
In the process, we were also able to reproduce reasonable results
for the mass spectrum of ground and excited states of charmonium
($c\overline{c}$) such as $\eta_c$, $\chi_{c0}$, and
$\small{J}/\psi$, which are found to be closer to data than the
mass spectrum we had found in earlier
 works of equal mass quarkonia \cite{hluf16,
bhatnagar18}. We then obtained the numerical values of leptonic
decay constants for these pseudoscalar and vector heavy-light
quarkonia with the same set of input parameters fixed from the mass spectrum. \\

All numerical calculations have been done using Matlab. We
selected the best set of 8 input parameters that gave good
matching with data for masses of ground and excited states of
heavy-light scalar, pseudoscalar, and vector quarkonia. This input
parameter set was found to be $C_0$= 0.139, $\omega_0$= 0.125 GeV,
$\Lambda_{QCD}$= 0.200 GeV, and $A_0$= 0.01, along with the input
quark masses $m_u$= 0.180 GeV, $m_s$= 0.350 GeV, $m_c$= 1.490 GeV,
and $m_b$= 5.070 GeV. The perturbation parameter $\gamma$, which
has been introduced to make the linear and harmonic terms of the
interaction kernel dimensionally consistent, is chosen to have the
form $\gamma=\displaystyle\frac{\omega_0^4}{C_0\beta^2}(3l+1)$
(where $\beta^2=\beta_{S,P,V}^2$) in order to produce reasonable
non-degenerate masses of 2S and 1D, 3S and 2D, 4S and 3D, etc.
states of heavy-light quarkonia. We have thus obtained results of
masses for $B_c$, $B_s$, $B$, $D_s$, $D$, and $\eta_c$,
$\chi_{c0}$, $\small{J}/\psi$, and are in reasonable agreement
with experimental data and other models. We will be using the
analytical forms of eigen functions for ground and excited states
of heavy-light quarkonia to evaluate the various transition
processes involving heavy-light scalar, pseudoscalar, and vector
quarkonia as future work.

}
\normalsize{

}
\end{document}